%% file: main.tex
\documentclass[conference]{IEEEtran}
\ifCLASSINFOpdf
\else
\fi
\usepackage{listings}
\usepackage{booktabs}
\usepackage{threeparttable}
\usepackage{hyperref}
\usepackage[htt]{hyphenat}
\usepackage[british]{babel}
\usepackage{xcolor}
\usepackage[figuresright]{rotating}
\hypersetup{
    colorlinks=true,
}

\definecolor{mGreen}{rgb}{0,0.6,0}
\definecolor{mGray}{rgb}{0.5,0.5,0.5}
\definecolor{mPurple}{rgb}{0.58,0,0.82}
\definecolor{backgroundColour}{rgb}{0.95,0.95,0.92}

\lstdefinestyle{CStyle}{
    commentstyle=\color{mGreen},
    keywordstyle=\color{magenta},
    stringstyle=\color{mPurple},
    basicstyle=\scriptsize\ttfamily,
    breaklines=true,
    captionpos=b,
    keepspaces=true,
    otherkeywords={status_t, ArrayObject,uint32_t, &},
    numbers=left,
    numbersep=1pt,
    showspaces=false,
    showstringspaces=false,
    showtabs=false,
    tabsize=2,
    language=C
}

\newcommand{\gl}[1]{\textcolor{red}{#1}}

\newcommand{\approachemph}{{\emph{para-rehosting}\xspace}}
\newcommand{\approach}{{para-rehosting}\xspace}
\newcommand{\Approach}{{Para-rehosting}\xspace}

\newcommand{\sys}{{\textit{PMCU}}\xspace}
\newcommand{\prefix}{{PMCU\_}}

\newcommand{\eg}{e.g.,}

\renewcommand{\paragraph}[1]{\vspace{5pt}\noindent\textbf{#1}}

\usepackage{adjustbox}

\usepackage{xspace}

\newcommand{\cid}{\textit{Type \uppercase\expandafter{\romannumeral1}}\xspace}
\newcommand{\hid}{\textit{Type \uppercase\expandafter{\romannumeral2}}\xspace}

\usepackage{multirow}
\usepackage{amssymb}
\usepackage{hyperref}
\makeatletter
\def\UrlAlphabet{%
      \do\a\do\b\do\c\do\d\do\e\do\f\do\g\do\h\do\i\do\j%
      \do\k\do\l\do\m\do\n\do\o\do\p\do\q\do\r\do\s\do\t%
      \do\u\do\v\do\w\do\x\do\y\do\z\do\A\do\B\do\C\do\D%
      \do\E\do\F\do\G\do\H\do\I\do\J\do\K\do\L\do\M\do\N%
      \do\O\do\P\do\Q\do\R\do\S\do\T\do\U\do\V\do\W\do\X%
      \do\Y\do\Z}
\def\UrlDigits{\do\1\do\2\do\3\do\4\do\5\do\6\do\7\do\8\do\9\do\0}
\g@addto@macro{\UrlBreaks}{\UrlOrds}
\g@addto@macro{\UrlBreaks}{\UrlAlphabet}
\g@addto@macro{\UrlBreaks}{\UrlDigits}
\PassOptionsToPackage{obeyspaces}{url}

\usepackage{booktabs,longtable}
\usepackage[center]{caption}
\usepackage{comment}

\begin{document}
%
\title{From Library Portability to Para-rehosting: \\Natively Executing Microcontroller Software \\on Commodity Hardware}

\author{
\IEEEauthorblockN{\IEEEauthorrefmark{2}\IEEEauthorrefmark{3}\IEEEauthorrefmark{4}\IEEEauthorrefmark{6}Wenqiang Li$^*$\thanks{$^*$Work was primarily done while visiting the University of Georgia.}, \IEEEauthorrefmark{3}Le Guan, 
\IEEEauthorrefmark{5}Jingqiang Lin, \IEEEauthorrefmark{3}Jiameng Shi, \IEEEauthorrefmark{6}Fengjun Li}
\IEEEauthorblockA{
\IEEEauthorrefmark{2}State Key Laboratory of Information Security, Institute of Information Engineering, Chinese Academy of Sciences \\
\IEEEauthorrefmark{3}Department of Computer Science, the University of Georgia, USA \\
\IEEEauthorrefmark{4}School of Cyber Security, University of Chinese Academy of Sciences \\
\IEEEauthorrefmark{5}School of Cyber Security, University of Science and Technology of China \\
\IEEEauthorrefmark{6}Department of Electrical Engineering and Computer Science, the University of Kansas, USA \\
liwenqiang@iie.ac.cn, \{leguan, jiameng\}@uga.edu, linjq@ustc.edu.cn, fli@ku.edu}
}


%


\IEEEoverridecommandlockouts
\makeatletter\def\@IEEEpubidpullup{6.5\baselineskip}\makeatother
\IEEEpubid{\parbox{\columnwidth}{
    Network and Distributed Systems Security (NDSS) Symposium 2021\\
    21-24 February 2021\\
    ISBN 1-891562-66-5\\
    https://dx.doi.org/10.14722/ndss.2021.24308\\
    www.ndss-symposium.org
}
\hspace{\columnsep}\makebox[\columnwidth]{}}

\maketitle

\begin{abstract}
Finding bugs in microcontroller (MCU) firmware is challenging, even
for device manufacturers who own the source code.
The MCU runs
different instruction sets than x86 and 
exposes a very different development environment. 
This invalidates many existing sophisticated software testing
tools on x86. 
To maintain a unified developing and testing environment,
a straightforward way is to re-compile the source code
into the native executable for a commodity machine (called rehosting).
However, ad-hoc re-hosting is
a daunting and tedious task and subject to many issues (library-dependence, kernel-dependence
and hardware-dependence). In this work, we systematically
explore the portability problem of MCU software and propose \approach to ease the
porting process. Specifically, we abstract and implement a portable MCU (\sys)
using the POSIX interface. It models common functions of
the MCU cores. For peripheral specific logic, we propose HAL-based peripheral function
replacement, in which high-level hardware functions are replaced with an
equivalent backend driver on the host. These backend drivers are
invoked by well-designed para-APIs and can be reused
across many MCU OSs. 
We categorize common HAL functions into four types and implement templates 
for quick backend development.
Using the proposed approach, we have successfully rehosted nine MCU OSs
including the widely deployed Amazon FreeRTOS, ARM Mbed OS, Zephyr and LiteOS.
To demonstrate the superiority of our approach in terms of security testing, 
we used off-the-shelf dynamic
analysis tools (AFL and ASAN) against the rehosted programs and discovered 28
previously-unknown bugs, among which 5 were confirmed by CVE and
the other 19 were confirmed by vendors at the time of writing.
\end{abstract}



\input{tex/1-intro}
\input{tex/2-bg}

\input{tex/3-abs}

\input{tex/4-mcu}
\input{tex/5-eval}
\input{tex/6-Disc}

\input{tex/8-avl}
\input{tex/7-con}
\section*{Acknowledgment}

We would like to thank the anonymous reviewers and our shepherd Dave (Jing)
Tian for constructive comments and feedback. The work reported in this paper
was supported in part by JFSG from the University of Georgia Research Foundation, Inc.,
NSF IIS-2014552, DGE-1565570, NSA Science of Security Initiative H98230-18-D-0009
and the Ripple University Blockchain Research Initiative.



%



\bibliographystyle{IEEEtranS}  
\bibliography{bib/rtos}

\appendices
\input{tex/app.tex}

\end{document}

%% file: tex/1-intro.tex
\section{Introduction}



It is commonly believed that the Internet of Things (IoT) is 
emerging as the third wave in the development of the Internet. 
By the year 2020, the world will have 50 billion connected
devices~\cite{evans2011internet}.
Among them, microcontroller units (MCUs) make up the majority.
It is projected that by 2020, there will be more than
35 billion MCU shipments~\cite{mcushipment}.
These small, resource-constraint devices 
are enabling ubiquitous connections and have changed
virtually every aspect of our lives. 
However, all of these benefits and conveniences come with broader and acute
security concerns. IoT devices are connected into the Internet, which directly exposes them to attackers.
Since these devices process and contain confidential data such as people’s health data, home
surveillance video, company secrets, etc., once compromised, devastating consequences can happen.
For example, 
FreeRTOS~\cite{freertos}, the leading operating system (OS) for MCUs,  was reported that its
$13$ critical vulnerabilities put a wide range of devices at risk of
compromise~\cite{bugs_of_tcpip}. In July 2019, \textit{urgent11}, a set of 11
vulnerabilities hit VxWorks, another popular MCU OS.
Once exploited, these bugs allow for remote code execution~\cite{urgent11}. 

In-house software testing is crucial for the security of the MCU software ecosystem. However, the development and test environment for MCU devices
is very different from that for commodity hardware.
Notably, the programs are cross-compiled and downloaded to the target device by
flashing the on-chip ROM.
To debug a program, a hardware dongle called in-circuit emulator (ICE)
is used to bridge the target device with the host via the JTAG interface.
The host machine then starts a GDBServer daemon that interprets
the ICE-specific control commands to the commands understandable by GDB.
Software testing of MCU software highly depends on the debugging features
integrated in the chip. For example, in ARM Cortex-M based MCUs,
the Data Watch and Trace (DWT) unit~\cite{dwt} can be used to profile the firmware
execution, and the Embedded Trace Macrocell (ETM) unit~\cite{etmv3} can be used to collect the execution trace.
However, it is clear that standard software testing
tools such as Valgrind~\cite{nethercote2007valgrind},  AddressSanitizer
(ASAN)~\cite{asan} cannot be supported due to the very different run-time environment.
We argue that the tools designed for x86 programs are more sophisticated
and superior to those designed for MCU firmware;
indeed, binary rewriting and instrumentation provide us with unprecedented 
insights to the program execution, which enables us to find software problems more efficiently.

Based on this observation, this work explores the idea of re-hosting MCU software and running them 
natively on commodity hardware. 
There are at least three benefits of this approach.
First, off-the-shelf dynamic analysis
tools for x86 can be readily used out-of-the-box.
Second, compared with running
on real MCU devices, multiple instances can be launched simultaneously, allowing
for a scalable and paralleled analysis. 
Third, the commodity hardware is much more powerful than MCU devices.
We can execute more testcases within a time period.

Intuitively,
with some manual efforts, we could possibly port a particular library for MCU to
the host machine and efficiently analyze its security. However, this is ad-hoc,
inaccurate, and sometimes extremely difficult. In particular, 1) the libc used in
MCU toolchains (such as newlib and redlib) has different designs compared with a full
fledged libc implementation such as the GNU libc. For example, the ARM Mbed OS
makes use of the function \texttt{software\_init\_hook()} to perform some
target-specific early initialization, which is not defined in the GNU libc;
2) More importantly, a single library is sometimes mingled with a set of supporting
OS routines, which must also be ported;
3) To make things worse, these routines
are subject to scheduling. Without considering the invocation sequence,
the intended logic can be easily violated;
4) If the
rehosted code needs to access real hardware on MCU, the behavior on the host becomes unpredictable.
All of these make ad-hoc porting a daunting task.


We propose \approachemph, a new technique aiming at making
re-hosting of MCU software to commodity hardware smoother. 
Different
from ad-hoc library porting, we support porting the MCU OS core entirely and allow for
incremental plug-and-play library porting.
Specifically, we abstract the whole machine model that most MCU
OSs follow and implement it (named \sys) with the \textit{Portable Operating
System Interface} (POSIX). Through a thin OS-specific glue layer, \sys can 
be compiled with upper-layer source code (including libraries and tasks) into a normal user-space program on the host.
\sys accurately models the common behaviors of real MCUs. As such, basic OS primitives including
scheduling, preemption, atomicity, etc. can be supported automatically. 

\sys also models the memory layout of a real firmware image.
An MCU often has a fixed memory map. MCU OSs correspondingly provide a linker
script to control how a memory region should be mapped into the physical
address space. To model the memory layout, we developed a template linker
script based on a popular MCU. It essentially squeezes the whole run-time
memory into a continuous memory region in the address space of a host process.

To support hardware-specific peripherals,
such as UART, Ethernet, SD card and CRC,
we
propose HAL-based peripheral function replacement.
An HAL layer allows upper-layer OS libraries to
interact with the hardware device at an abstract level rather than at the
hardware level. We identify high-level HAL functions and replace them
with equivalent handlers on the host.
This simplifies the porting effort as well as improves I/O performance. Per HAL
function, a set of para-APIs are defined for the HAL library to invoke.
Correspondingly, backend drivers are implemented on the host.
It can be shared among multiple HALs from different vendors.
In this
sense, our design follows the spirit of para-virtualization in which the guest
OS has to be explicitly modified to benefit from optimized I/O implementation in the backend (thus the name \approach).
To speedup backend development for new peripherals,
we categorize common HAL functions into four types (IO, 
storage, computing accelerator, and dummy) and implement the corresponding templates.
Developers only need to figure out the appropriate peripheral categorization
and quickly customize the implementation.

We have evaluated our approach against nine MCU OSs
including the widely deployed Amazon FreeRTOS, ARM Mbed OS, Zephyr, LiteOS, etc.
We successfully compiled and executed 84.21\%, 76.47\%, 89.47\% and 63.64\% of
all the libraries shipped with FreeRTOS, ARM Mbed OS, Zephyr and LiteOS respectively.
Moreover, our HAL backends support most peripherals
in the SDKs of the NXP and STMicroelectronics devices.
To demonstrate the superiority of our approach in terms of  security  testing,
we further leveraged
AFL~\cite{afl} and ASAN~\cite{asan} to test several popular libraries in the
ecosystems of Amazon FreeRTOS, ARM Mbed OS, Zephyr and LiteOS.
Our tool has helped us find 28 previously-unknown bugs.
We have responsibly reported all of them to the influenced vendors.
At the time of writing, 5 were confirmed by CVE and the other 19 were confirmed by vendors.

In summary, we made the following contributions.

\begin{itemize}
	
	\item We proposed \approach to natively port MCU software on commodity
	hardware. It accurately models common MCU behaviors and depends on para-APIs
	to support HAL-enabled hardware logic.

	
	\item We implemented the idea of \approach by prototyping a portable MCU based on POSIX and
	a set
	of para-APIs (and the backends).
	Our prototype supports nine MCU OSs, including FreeRTOS, ARM Mbed OS, Zephyr,
	and LiteOS currently.
	
	
	\item We used fuzz testing to analyze several popular libraries used in MCU
	and identified 28 previously-unknown bugs. 
	Our tool is open source available at \url{https://github.com/MCUSec/para-rehosting}.
\end{itemize}

%% file: tex/2-bg.tex
\section{Background}

\subsection{Microcontroller Units}



Until recent years,
MCUs were considered as specialized computer
systems that are embedded into some other devices, as contrary to 
general-purpose commodity computing systems such as personal computers (PCs)
or mobile devices.
With the emergence of IoT, now
MCUs have been tasked with more diverse missions and are
at the center of many of the innovations in the cost- and power-efficient
IoT space.
Examples include 
home automation, wearable devices, smart city, smart manufacturing, etc.~\cite{microcontroller}. 

MCUs have evolved from 8-bit design to 16-bit design. 
Now, 32-bit MCUs have dominated the market,
accounting for 55\% of the total MCU sales~\cite{matas2013mcu}.
In the MCU segment, the major players include
the ARM Cortex-M family MCUs, MIPS MCUs, and Atmel AVR, etc.
To keep energy-efficient, MCUs are equipped with
limited computing and memory resources.
For example, the state-of-the-art ARM Cortex-M4 processor
often runs at a frequency of around 100 MHz and the
size of SRAM is about 200 KB.

From the viewpoint of a programmer, the most
remarkable difference between PC/mobile processors and MCUs is that
MCUs do not support MMU. 
As a result, the application code and the OS kernel code 
have to be mingled together in a flat memory address space.
We call the resulting executable binary as a firmware image.
Without virtual memory support, the Linux kernel cannot run
on top of MCUs.
Another characteristic of MCUs is that they
are highly heterogeneous.
Each MCU could support a distinct set of peripherals.
The peripherals could be custom-made and thus have different specifications.

\subsection{Software Ecosystem for MCUs} \label{sec:MCUOS}
Due to the lack of the MMU support, traditional OSs such as Linux cannot
run atop MCUs.
Instead, since MCUs have a long history
of being used in safety-critical real-time applications, 
many \textit{real-time operating systems} (RTOSs) have been developed for them.
Given that MCUs have become the driving hardware for the emerging IoT technology,
big companies have quickly begun to invest on building their ecosystems for MCU devices.
FreeRTOS~\cite{freertos}, arguably the most widely deployed RTOS for MCUs,
has been acquired by Amazon in 2017.
As an essential part of the Amazon Web Service (AWS),
FreeRTOS has been extended with libraries that enable local and cloud
connectivity, security, and over-the-air (OTA) updates.
ARM Mbed OS~\cite{mbedos}, backed by ARM, is another IoT RTOS dedicated for
ARM MCU devices.
It includes all the needed features to develop a connected product,
including security, 
connectivity, an RTOS, and drivers for sensors and I/O devices.
Zephyr~\cite{zephyr}, backed by the Linux Foundation,
is an IoT RTOS that integrates all the necessary components and libraries
required to develop a full application.
Finally, LiteOS~\cite{liteos}, backed by Huawei,
is a lightweight IoT RTOS that receives wide adoption in China.
Based on a recent IoT developer survey~\cite{rtosmarket},
FreeRTOS continues to be the dominating RTOS for constrained IoT devices,
while ARM Mbed OS, Zephyr and LiteOS are creating successful open source communities.

\begin{figure}[t]
\centering
\includegraphics[width=\columnwidth]{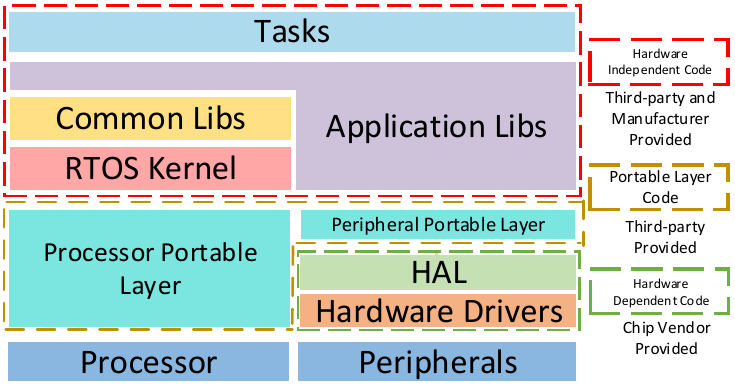}%
\caption{FreeRTOS software stack}
\label{fig:stack}
\end{figure}

\paragraph{Software Stack.}
It is quite clear from the aforementioned facts that big companies 
are developing their own IoT ecosystems.
They aim to create a smooth and pleasant experience for developers 
by providing a comprehensive software stack and an active community.
Eventually, the developers would stick to their ecosystem.
In what follows, we briefly introduce the software stack of the FreeRTOS as shown in Figure~\ref{fig:stack}.
Other RTOS ecosystems follow a very similar design.

Amazon FreeRTOS provides both a RTOS kernel as well as many libraries/middleware
that make it easy to securely connect to the cloud or other devices.
At the core is a RTOS kernel which is responsible for task management,
scheduling, memory management, interrupt management, message delivery, etc.
Task is the basic scheduling unit in a RTOS, similar to a thread in the Linux OS.
Different from Linux threads, all tasks in the RTOS share the same address space with kernel.
There are several built-in tasks that are automatically started by the kernel.
They are mainly used for system-level event management. 

There are three types of libraries, namely common libraries, application libraries and portable layer libraries.
Common libraries extend the kernel functionality with additional data structures and functions,
such as atomic operations.
Application libraries are standalone libraries for connectivity and remote management,
such as MQTT and device shadow.
Typically, an application library also acts as a
dedicated system task, serving for other tasks.
Lastly, the portable layer libraries handle device specifics.
They serve as an adaptor layer for processors (e.g., ARM and MIPS)
and peripherals (e.g., network).
Many ports for different hardware
have been provided officially or unofficially.
In developing a port, developers are only responsible for implementing
a number of downstream APIs
that the upper-layer libraries rely on.
Conversely, these downstream APIs have to invoke certain upstream APIs provided
by the RTOS kernel to fulfill their functionality.

\paragraph{Hardware Abstraction Layer.}
Another important piece of software in MCU firmware
is provided chip vendors.
For example, STM32 provides each kind of chip with an SDK that
includes low-level hardware
drivers as well as a \textit{hardware abstraction layer} (HAL)~\cite{stm32cubel5}.
An HAL layer acts as a bridge between software and hardware.
It allows upper-layer libraries to interact with a hardware device
at a general and abstract level rather than at the hardware level.
Therefore, it is invoked by the portable layer to interact with
the peripherals.

%% file: tex/3-abs.tex
\section{Overview}

Our work explores the portability problem of open-source MCU OSs for the purpose of
finding software bugs in them. We first present a motivating example. Then we
give an overview of the proposed \approach technique.


\subsection{Motivating Example}

There has been a bunch of advanced \emph{dynamic} analysis techniques
to test the security of software, such as 
fuzz testing~\cite{afl}, memory checker~\cite{asan},
and dynamic instrumentation~\cite{pin}.
In practice, a prerequisite of using these tools is that the target binary
must be executable on a PC.
If we attempt to port a particular MCU library, taking
the MQTT library of FreeRTOS as an example,
many problems arise.

\begin{lstlisting}[style=CStyle, numbers=none]
void IotMqtt_ReceiveCallback( ... )
{   ...
	_getIncomingPacket( pNetworkConnection, ..., &incomingPacket );
}
\end{lstlisting}

This function is invoked whenever the MQTT task receives data from the
network (recall that MQTT is an application library). The function \texttt{\_getIncomingPacket} actually fetches the data
from the network connection maintained in
\texttt{pMqttConnection->pNetworkInterface}. More specifically, the
\texttt{NetworkInterface} is also manipulated by another task called
\texttt{IP-task}. The MQTT task and the \texttt{IP-task} synchronize with each other through
the \texttt{Queue} mechanism defined by the FreeRTOS kernel.
To port the MQTT
library, it becomes essential to also port the \texttt{IP-task}
(\textbf{library-dependence}) and the kernel (\textbf{kernel-dependence}). In
the \texttt{IP-task}, if we continue to track down the source code, it will eventually
call the MAC driver level function \texttt{SPI\_WIFI\_ReceiveData()} (on the STM32f756 chip), which
in turn uses the SPI protocol to transfer data. Under the hook, it  receives
an amount of data in no-blocking mode with interrupt. To enable interrupt, the
following statement is invoked.

\begin{lstlisting}[style=CStyle, numbers=none]
__HAL_SPI_ENABLE_IT(hspi, (SPI_IT_RXNE | SPI_IT_ERR));
\end{lstlisting}

The macros is actually defined as 

\begin{lstlisting}[style=CStyle, numbers=none]
#define __HAL_SPI_ENABLE_IT(__HANDLE__, __INTERRUPT__) SET_BIT((__HANDLE__)->Instance->CR2, (__INTERRUPT__))
\end{lstlisting}

It sets certain bits of a hardware register at a fix address to enable interrupt.
Na\"ively compiling this code will lead to unpredictable behaviors or even crash the program
(\textbf{hardware-dependence}). Due to the aforementioned
\textbf{library-dependence}, \textbf{kernel-dependence} and
\textbf{hardware-dependence} issues, porting an MCU library
to another host is regarded
as a very daunting and tedious task.





\begin{figure}[t]
\centering
\includegraphics[width=\columnwidth]{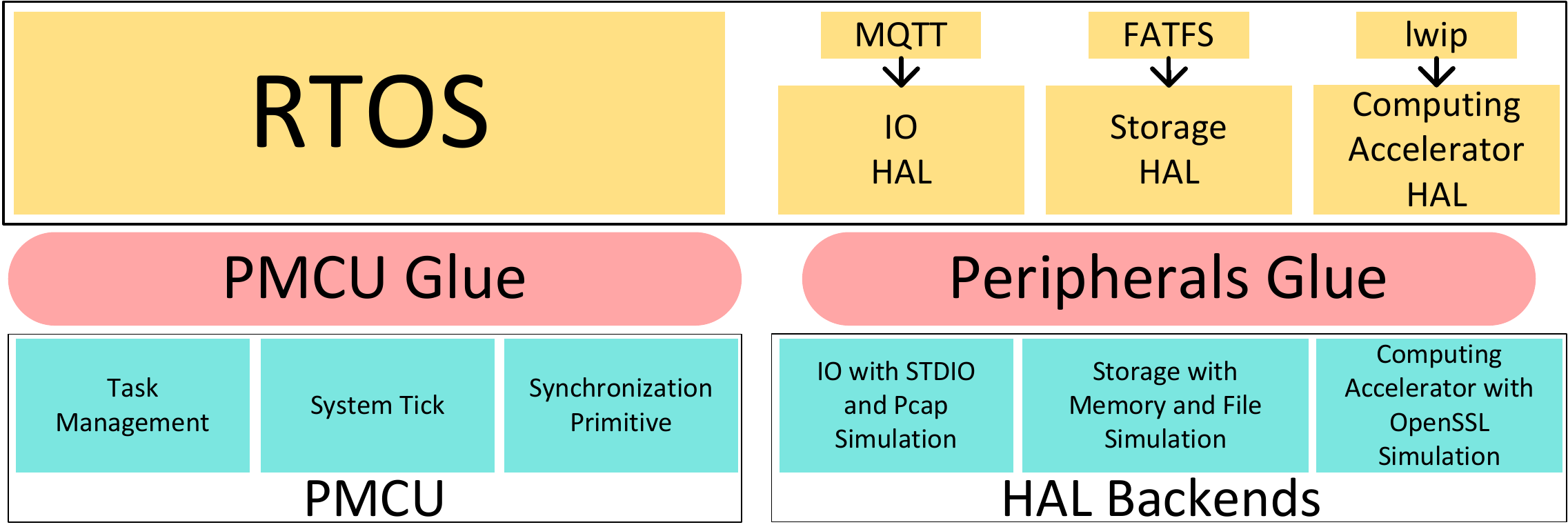}%

\caption{Para-rehosting overview}
\label{fig:components}
\end{figure}

\subsection{Para-rehosting}

We propose para-rehosting to ease the complexity of porting MCU software.
In para-rehosting, 
we provide common backend implementations in the host machine
that simulate processor and peripheral behaviors.
The upper-layer software only needs slight modifications
to accommodate the backend.
With para-rehosting,
we can compile the
whole RTOS logic, including the kernel, built-in tasks, normal tasks
altogether into a single host program.
They share the same virtual address space of
a process on the host, just as they share the same flat physical address space
on a real MCU.

As shown in Figure~\ref{fig:components}, the proposed system is comprised of
two major components. The portable MCU (\sys) is a host backend that
models the common functions of
an MCU and the available memory resources on it. It can be easily ported to different
MCU OSs. Specifically, \sys simulates task scheduling and the system tick. It
also provides basic synchronization primitives to MCU OSs. All of these
functions are essential for an MCU OS. 
Our implementation is modularized. 
Each abstract module
is placed in a separate C source file and included in the project during compilation on demand.
For each supported MCU OS, there is a thin glue layer to
accommodate specific OS requirements.
Moreover, we use a linker script to
manage the layout of the resulting program so that it resembles that on a real
device. With \sys, we can accurately re-host an MCU firmware that does
not rely on any hardware-specific peripherals.

The other component, called HAL-based peripheral function replacement,
handles hardware-specific peripheral logic. 
As such, it addresses the hardware-dependence issue.
Since HALs abstract away low-level hardware
details from the programmer, we can easily replace its high-level function
with an equivalent handler. 
Note that the HALs for devices from different vendors
are typically different. 
We cannot simply implement a high-level replacement and use it for all the
devices. We solve this problem by implementing the HAL function’s semantics
as a common backend, and require some manual work for each HAL library to
invoke the backend drivers.  Correspondingly, the parameters need to be
adjusted and return values need to be captured properly.

%% file: tex/4-mcu.tex
\section{Portable MCU}
\label{sec:abstracing}

Portable MCU (\sys) models the common functions of an MCU and the available memory resources on it.
Our prototype uses the widely adopted
POSIX interface to simulate the abstract common functions.
In this way, the rehosted firmware can be executed and analyzed
in all UNIX-like systems (including Linux).
In Table~\ref{case_study}, we summarize the needed changes
to port \sys to popular MCU OSs.
It includes the upstream functions that portable layer libraries
rely on, the downstream functions that they
provide, and the common backend functions that they invoke
to actually simulate the abstract MCU.





\begin{figure}[t]
\centering
\includegraphics[width=\columnwidth]{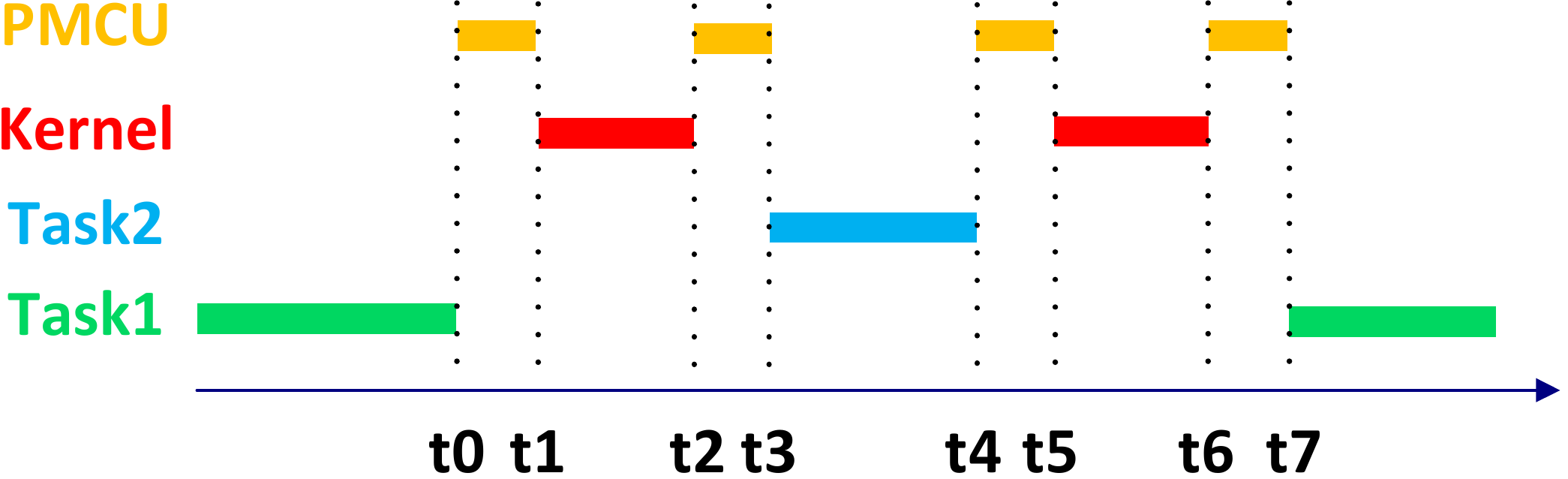}%
\caption{MCU OS execution with two tasks}
\label{fig:execution}
\end{figure}

\subsection{Abstracting Machine Functions}

\subsubsection{Task Management} 

Different from traditional MCU systems in which a single task monopolizes the
processor, in the IoT era, a variety of tasks run simultaneously. This
necessitates a multi-programming environment. Multi-task support has become a
standard feature in major MCU OSs, which \sys needs to simulate
correspondingly.

We use a process in Linux to simulate the entire MCU firmware,
and a thread to simulate a task.
The thread-process model of Linux provides the basis
for simulating the task-firmware model of MCU.
In particular, in Linux, all threads share
the same virtual memory space of a process,
and have their own stacks.
In MCU firmware, all tasks share
the same physical memory space of a MCU,
and have their own stacks.
In \sys, when a task is created by the MCU OS, a \texttt{pthread} is created,
with the \texttt{start\_routine} initialized to be the starting function of the task.
Task switch is also supported transparently because the \texttt{pthread} library
is responsible for saving and restoring the context.
In most MCU OSs, at a particular time, only one thread can be executed due
to the immature multi-core support~\cite{freertossmp,mbedsmp}.
As a result, we need to ensure that a thread exclusively
occupies the current process.
That is, although conceptually there are multiple threads, only one is runnable.
This is achieved by using a combination of the \texttt{signal} and \texttt{mutex} mechanisms
in POSIX.
Only the thread holding the mutex is allowed to execute.
When the current thread yields or is preempted,
it unlocks the mutex to allow others to execute
and suspends itself by calling \texttt{sig\_wait()}.
To resume, it must receive a signal \texttt{SIGUSR1} and grab the same mutex in the
corresponding signal handler.

By default, when creating a thread,  the \texttt{pthread} library allocates a
memory region used for stack, which is out of the control. In
Section~\ref{sec:linkerscript}, we explain how to make sure the location of a task
stack is within the memory map of a real MCU.


\subsubsection{System Tick} 
\label{3_system_tick}

The MCU's time is shared among multiple tasks simultaneously. A timer, called
system tick is usually used to invoke the scheduler to switch tasks
periodically.
To support preemption, the RTOS kernel is periodically awakened by a system
timer, which is often implemented by a counter hardware in a real MCU.
Naturally, we leverage the system timer in POSIX to periodically raise a timer
signal to simulate this process. In particular, we use \texttt{setitimer()} to
deliver a \texttt{SIGVTALRM} signal of type \texttt{ITIMER\_VIRTUAL} periodically.
\texttt{ITIMER\_VIRTUAL} timer counts down against the user-mode CPU time
consumed by all threads in the process. As such, it avoids issues caused by 
unpredictable time allocated to a process. Inside the \texttt{SIGVTALRM}
signal handler, \sys invokes the scheduler to resume the next task. Note that
the scheduling algorithm and task priority are totally transparent to \sys.
Note that due to the unpredictable behavior of the timer on the Linux machine,
we cannot guarantee that each simulation yields exactly the same execution
path. This is also the case on a real device. We demonstrate that this
nondeterministic behavior does not influence bugs finding in practice later.

\subsubsection{Synchronization Primitive}

Synchronization is a basic OS primitive. Once violated, the execution is
subject to crashes. For example, when a task enters the critical section, some
RTOSs need to disable interrupts (including the system tick). If \sys allows
system tick (and thus task scheduling) in a critical section, race condition
may happen. \sys simulates disabling interrupt by keeping track of the current
interrupt status in a global variable \texttt{\prefix INT\_ENABLED}. When a
\texttt{SIGVTALRM} signal occurs, if \texttt{\prefix INT\_ENABLED} is cleared,
the handler returns immediately with the \texttt{\prefix pending} variable 
set which indicates the scheduler should be called after exiting the critical section. Otherwise, the handler performs a normal
scheduling process. Additionally, \sys ensures that it does not interfere
with the RTOS kernel itself, since \sys also needs to access critical RTOS data
structures.


\subsubsection{A Running Example}
In Figure~\ref{fig:execution}, we showcase a running example for an MCU OS
execution with two task. \texttt{Task1} is executing until \texttt{t0} when a
\texttt{SIGVTALRM} signal occurs requesting for a potential task switch. The
handler then kicks in (represented by \sys) to invoke the systick handler in
the kernel (\texttt{t0-t1}). The handler in the kernel selects the next task
and sends a signal \texttt{SIGUSR1} (\texttt{t1-t2}). The signal handler
simply schedules the thread corresponding to the selected task
(\texttt{Task2}) to be runnable (\texttt{t2-t3}). As a result, \texttt{Task2}
begins to execute starting from \texttt{t3}. The execution sequence from
\texttt{t4} to \texttt{t7} is self-explanatory.

\subsection{Memory Layout Enforcement}

\label{sec:linkerscript}

As mentioned before, the memory layout of the rehosted program (including
code, data, stack, bss, etc.) is different from that in a real device. 
For example, the default linker on a Linux machine places the code segments
starting from 0x08048000 on x86-32 and from 0x400000 on x86-64 (if PIE is not enabled), while the
linker for an MCU compiler typically places the code segment starting from
zero. 
This is controlled by a file called linker script. Typically, an MCU has
a flash memory starting from zero and an separate SRAM at another offset. To
be able to more accurately capture the program misbehavior caused by memory
errors, we need to place the code and data segments following the memory map
in the MCU. We could simply reuse the linker script available for an MCU
firmware, however, three problems arise.

First, the stack of a Linux program is allocated by the kernel rather than
based on the linker script. Specifically, the stack grows downwards from
0x7FFFFFFF on x86-32 and 0x7fbFFFFFFF on x86-64 (without regard to
randomization). Moreover, the default stack size of a thread is limited to 2 MB. This number
is even beyond the total SRAM size on many MCUs.  To solve this problem,
we implemented a trampoline function
called \texttt{stack\_switching(void *newSP, void *newFun)} in assembly that
explicitly manipulates the \texttt{SP} register so that the thread uses a
newly specified stack. It then jumps to the function \texttt{newFun()}. 
We wrapped the start routine of new tasks (say \texttt{start\_routine()}) into 
\texttt{stack\_switching(newSP,start\_routine)},
so that the function switches the
stack to the one allocated based on the MCU linker script and then jumps to
the real start function \texttt{start\_routine()}. We note that
similar results can be achieved by using the \texttt{makecontext/setcontext} APIs.

Second, the heap management in MCU OSs conflicts
with that in Linux. 
The MCU libc allocator does not implement the underlying
\texttt{\_sbk()} but relies on the developer
to provide a device-specific implementation according to the heap range specified in the linker script.
Therefore, the provided \texttt{\_sbk()} conflicts with
that in the Linux libc allocator.
We have two choices here.
First, we could suppress the Linux allocator. In this way,
we can precisely simulate the heap layout as specified in the
linker script.
Second, we could suppress the firmware-provided \texttt{\_sbk()} and use the Linux version. In this way,
we sacrifice the accurate emulation of memory layout.
This might cause problems in bug re-production and exploitation.
However, using the Linux allocator allows
us to harvest the bug-finding capability
of ASAN out-of-the-box. 

Some MCU OSs (such as FreeRTOS and LiteOS) provide their own allocators. For them, we also have two strategies; 1)
we can replace their heap implementation with that on Linux, but sacrifice the coverage of customized allocators;
2) we can keep their implementations, but sacrifice ASAN in finding memory errors in heap.

Finally, the entry point of an MCU program is the function
\texttt{Reset\_Hander()}, rather than \texttt{\_\_start()} in a regular Linux
program. 
\texttt{\_\_start()} performs necessary
initialization of the execution environment and jumps to the main
function, while the \texttt{Reset\_Hander()} initializes the simulated memory
on the  memory map of the real device, such as  coping the data sections from
the ``flash'' to the ``SRAM''. We keep the entry point of the rehosted program
as \texttt{\_\_start()} but explicitly invoke the modified
\texttt{Reset\_Hander()} before the main function is executed.

\section{HAL-based Peripheral Function Replacement}

Some peripheral functions (e.g., networking)
are indefensible for IoT applications.
If not supported, many MCU firmware cannot be rehosted.
To avoid handling low-level hardware details,  we leverage HALs supported by
major MCU vendors.
Specifically,  we replace the high-level HAL routines with
equivalent handlers on the host. These handlers are implemented per peripheral function.
We call them HAL backends in this paper and they are exposed to HAL
libraries as para-APIs. Inside the HAL library, we only need to identify the
corresponding routines and make minor modifications to invoke these para-APIs.
As such, a backend which corresponds to a peripheral function,
can be shared by multiple HALs.
To ease the process of developing backends for new peripheral functions,
we further categorize common peripheral HAL functions 
into four types (IO,  storage,  computing  accelerator,  and  dummy)
and implement templates for quick backend development.
In Table~\ref{peripheral_drivers},
we summarize the categorization and the supported peripherals
in each categorization.
For each peripheral function,
we also list the function names of frontends for two
HALs (STMicroelectronics and NXP) and function names of the backend.



\subsection{I/O Peripherals} 
I/O peripherals are generally used to communicate with the real-world.
This includes UART, I2C, SPI, Ethernet, etc.
In this work, we first develop a generic I/O backend
which transparently 
bridges the peripheral I/O to host-side standard I/O.
As such, the re-hosted firmware has the host console as
its I/O interface. This approach works well for simple peripherals  such
as UART, which is typically used for debugging output.
Another benefit of this approach is that \texttt{STDIN}
can be readily overridden by a fuzzer like AFL to feed
testcases to the firmware execution.



\paragraph{Network.}
We observe that the network function is one of the most
widely used I/O, but redirecting network traffic to the \texttt{STDIO} 
as is done in the generic backend
rarely fulfils our re-hosting purposes,
because the libraries using network would block 
due to the missing protocol simulation in the generic backend.
To simulate a network interface,
we also developed a high-fidelity network backend based on
the \textit{Packet Capture} (PACP)
library~\cite{jacobson1989tcpdump}.
It enables re-hosted firmware to actually access the Internet.
The most essential tasks of a network
driver are to initialize the \textit{network interface card} (NIC), send
out-going messages and receive in-going messages through the NIC. The PCAP
library enables easy access to all packets on a host and thus fulfills our
requirements.

Specifically, we provide three para-APIs 
\texttt{HAL\_BE\_NetworkInit()}, \texttt{HAL\_BE\_NetworkSend()} and \texttt{HAL\_BE\_NetworkReceive()}.
They are to be invoked by the relevant frontend
routines in the IP layer of the HAL libraries.
The function \texttt{HAL\_BE\_NetworkInit()} opens a live physical Ethernet interface
on the host machine using the PCAP API \texttt{pcap\_open\_live()}, 
which returns a handler associated with this NIC.
To send out a data packet, the para-API, \texttt{HAL\_BE\_NetworkSend()} 
extracts the packet buffer pointer and packet length
from the provided data structure and then directly 
invokes the PCAP API \texttt{pcap\_sendpacket()}
to output the packet to Ethernet.
To receive a message, the para-API \texttt{HAL\_BE\_NetworkReceive()} 
is used to call the blocking PCAP API \texttt{pcap\_dispatch()} to receive a packet.
The packet is reconstructed and transmitted to the upper layers of 
the MCU OS stack by calling the corresponding callback functions.

\subsection{Storage Peripherals}
Storage peripherals are generally used as 
hardware medium for file systems 
such as FAT or littlefs. Popular storage medium 
used in IoT devices includes MMC, NAND, NOR and SD.
For the HAL of these four types of storage devices,
we develop a generic storage backend which operate on the host file system.
We can safely abstract away the details of medium access characteristics.
Specifically, we use a file to store the whole file system of the firmware. 
To mount the medium, the para-API \texttt{HAL\_BE\_Storage\_Init()} is invoked.
It maps the the whole file contents into the memory
as the raw medium data.
Then, storage read/write operations are conducted
by invoking \texttt{HAL\_BE\_Storage\_read()} and \texttt{HAL\_BE\_Storage\_write()}, which
simulate medium access by reading/writing the memory.

\subsection{Computing Accelerator Peripherals}
Computing accelerator peripherals
provide hardware-assisted implementation of popular
algorithms,
including many cryptography algorithms and the random number generator.
We mainly used the OpenSSL library to simulate
these algorithms and feed the results to the frontend functions.

\subsection{Dummy Peripherals}
Dummy peripherals generally do not
perform actions that may
affect the execution of the firmware.
Therefore, we can safely return a success code to
the HAL frontend, or just a \texttt{void} if the function does not expect
a return value.
This includes PWR (Power Controller), RCC (Reset and Clock Control), ICACHE (Instruction Cache) and so on. 

%% file: tex/5-eval.tex
\section{Evaluation}


We have developed a para-rehosting prototype on a x86-based PC
running the Ubuntu 16.04 OS. 
Our prototype supports nine MCU OSs,
including Amazon FreeRTOS~\cite{amazonfreertos}, ARM Mbed OS~\cite{mbedos}, Zephyr~\cite{zephyr},
Huawei LiteOS~\cite{liteos}, Atomosher~\cite{atomosher}, brtos~\cite{brtos}, f9-kernel~\cite{f9}, FunkOS~\cite{funkos}, and 
TNeo~\cite{tneo}.
For HAL functions, our prototype covers most 
peripherals supported in the SDK of NXP and STMicroelectronics.
In total, \sys is comprised of 497 LoC, 
including 294 for task management, 165 for system tick and 38 for synchronization primitive, which are OS-agnostic. 
Less than 50 LoC are needed for the glue layer
of each supported MCU OS.
Dozens to hundreds of LoCs were developed for each HAL backend.
Modifications made to HAL libraries that invoke HAL backend are negligible (less than 30 LoC for each).
All the details have been reported 
in Table~\ref{case_study} for MCU OSs and
Table~\ref{peripheral_drivers} for HALs.



We evaluated \approach from four aspects.
First, we evaluated the rehosting capability of \approach.
This was conducted from two dimensions -- library support
for MCU ecosystems and peripheral support in HAL.
Second,
since the ultimate goal of this work is to enable security testing for MCU firmware,
we used an off-the-shelf fuzzing tool AFL~\cite{afl} to
test the firmware logic compiled by \approach,
and compared the results with other solutions.
Third, we demonstrated its bug finding capability and explained some vulnerabilities
disclosed with the help of the proposed system.
Finally, due to the architectural difference and re-compilation,
we designed an experiment to identify the gap of running the same 
firmware logic on a real device and with \approach.
Unless stated otherwise, all our experiments
were conducted on a PC with an Intel Core i7-8700 CPU and 8 GB DRAM.


\subsection{Rehosting Capability}

This section describe the rehosting capability of the proposed system.
As mentioned before, an MCU ecosystem often offers a comprehensive software stack
to attract developers and manufacturers.
We obtained the library information for four representative
MCU OSs including Amazon FreeRTOS, ARM Mbed OS, Zephyr and LiteOS
from the corresponding official documentation pages or Github repositories,
and counted the libraries supported by our prototype.
The results are summarized in Table~\ref{tab:libsupport_tab}.
Our prototype supports 84.21\%, 76.47\%, 89.47\% and 63.64\% of all the
libraries shipped with FreeRTOS, ARM Mbed OS, Zephyr and LiteOS respectively.

We also summarized the supported HAL functions for
two popular MCU chip vendors, STMicroelectronics and NXP.
This statistic indicates how many peripherals can be supported
by our prototype.
The results are summarized in Table~\ref{peripheral_drivers}.
In the table, we also list the relevant frontend
and backend functions for each peripheral.
It is worth mentioning that after gluing the source code with
\sys and HAL backends, we did not found any failed compilation.
Particularly, no architecture-specific assembly code was found
in MCU libraries. This is partially because hardware-neutral MCU libraries 
have been widely adopted in the MCU ecosystem.



\begin{table}[t]
  \centering
    \caption{Results of persistent mode fuzzing with ASAN enabled}
    \begin{adjustbox}{max width=\columnwidth}
    \begin{tabular}{l|lrrrr}
    \toprule
    \textbf{RTOS} & \textbf{Library} & \textbf{Speed (\#/sec)} &  \textbf{Crashes} & \textbf{Real Bugs} & \textbf{Total Paths} \\
    \midrule
    \multirow{8}[2]{*}{FreeRTOS} & TCP/IP & 4,568.33 & 78 & 3 & 230 \\
          & MQTTv1 & 5,622.56 & 32 & 1 & 536 \\
          & MQTTv2 & 2,754.39 & 18 & 1 & 387 \\
          & FATFS  & 1,516.95 & 1  & 1 & 1,502 \\
          & Tinycbor & 7,975.00 & 0 & 0 & 272 \\
          & Jsmn & 21,828.81 & 0 & 0 & 198 \\
          & uTasker Modbus & 664.90 & 26 & 5 & 79 \\
          & lwip (latest) & 1,294.56 & 9 & 2 & 158 \\
          & lwip (2.1.2) & 1,063.47 & 7 & 2 & 139 \\
    \midrule
    \multirow{4}[2]{*}{MbedOS} & MQTT  & 814.18 & 57 & 1 & 104 \\
          & CoAP Parser & 15,025.01 & 95 & 1 & 522 \\
          & CoAP Builder & 1,553.65 & 15 & 1 & 502 \\
          & Client-Cli & 1,131.15 & 103 & 2 & 435 \\
    \midrule
    Zephyr & MQTT  & 1,311.56 & 0 & 0 & 174 \\
    \midrule
    \multirow{2}[2]{*}{LiteOS} & MQTT  & 667.13  & 4 & 2 & 42 \\
                              & LWM2M & 10,352.67 & 23 & 2 & 243 \\
    \midrule
    Baremetal & STM-PLC  & 2,552.81 & 41 &9 & 323 \\
    \bottomrule
    \end{tabular}%
    \end{adjustbox}
    \label{tab:performance}%
\end{table}%

\subsection{Fuzzing Performance}\label{fuzzing_perf}

The ultimate purpose of our tool is to enable software testing and
help find vulnerabilities in MCU firmware.
We used AFL~\cite{afl} plus ASAN~\cite{asan}, one of the most efficient
dynamic analysis combinations to test several
libraries for each ported RTOS.
These tools can be used out-of-the-box.
In what follows, we show the fuzzing performance against the firmware logic
compiled with our tool. Then we compare its performance with existing solutions.

\subsubsection{Fuzzing Popular Libs with \approach}

We fuzzed 17 popular libraries for MCU across different OS ecosystems.
Three criteria were considered in selecting libraries.
First, the core logic of the library should not have been replaced by para-APIs.
Otherwise, we would have been fuzzing the para-APIs.
Second, the code needs to be easy to tame.
Otherwise, tremendous efforts are needed to understand and accommodate the code
to AFL.
Third, the library is popular.
With modest taming efforts,
we tested these libraries across the four RTOSs shown in Table~\ref{tab:performance}.
FreeRTOS+TCP is a TCP/IP implementation officially supported by FreeRTOS~\cite{freertostcp}.
In 2018, there were $13$ critical vulnerabilities reported in this library by Zlab~\cite{bugs_of_tcpip}. 
MQTT  protocol is one of the most popular connection protocol used in IoT devices.
It has been widely used in many commercial smart home solutions.
FATFS is an embedded FAT file system that has already been used in commercial products~\cite{fatfs}.
CBOR (Concise Binary Object Representation) is 
a binary data serialization format.
Jsmn is a world fastest JSON parser/tokenizer.
Modbus is a de facto standard communication protocol 
and is now a commonly available means of 
connecting industrial electronic devices.
lwIP (lightweight IP) is a widely used open-source 
TCP/IP stack designed for embedded systems.
CoAP is another popular IoT protocol which focuses on one-to-one communication.
Client-Cli is a command line library, mainly
used by the Mbed OS to parse the remote commands sent to the device.
LWM2M (LightWeight Machine-to-Machine) is a lightweight protocol suitable
for M2M or IoT device management and service enablement.
Finally, STM-PLC (STM Programmable Logic Controller) is a PLC SDK that
turns an STM32 MCU into an industrial PLC.
We built these libraries with ASAN enabled for improving bugs visibility.
All the results were obtained within one hour and with AFL's persistent mode on.

In Table~\ref{tab:performance}, we present the results.
Generally speaking,
the fuzzing throughput is in line with the commonly perceived number, although
variations were observed depending on the libraries. First, the throughputs of
fuzzing the CoAP parser, Jsmn and LWM2M libraries are substantially faster than others. This is
because these libraries only involve a single independent task that
analyzes data packets provided by the fuzzer. For the CoAP builder library,
although it also has only one task, it constructs the message based on the
data structure \texttt{sn\_coap\_hdr\_s}, which contains many pointers. The
library needs to dynamically allocate and free buffers for each pointer. Due
to the way ASAN handles allocation, heavy overhead was observed. 
For other libraries, more than one tasks are involved and these tasks extensively use
message queues to pass packets. This leads to multiple threads waiting for
each other.


In most libraries, we could find the first crash almost instantly.
But we have not found any bugs in the MQTT library of Zephyr
and Tinycbor and Jsmn library of FreeRTOS.
For Tinycbor and Jsmn, since these libraries are quite simple,
porting them without our approach is still manageable.
We suspect they have been heavily fuzzed.
For the MQTT library of Zephyr,
this may be partially attributed to the fact that
this library has also been intensively tested because Zephyr officially supports
a simulator that helps developers to quickly prototype their products~\cite{zephyremu}.
But note that their simulator is specifically designed for Zephyr.
We discuss more in Section~\ref{sec:nativecomparison}.

Last but not least, we were able to discover $9$ bugs out
of the $13$ bugs reported by Zlab~\cite{bugs_of_tcpip} in the FreeRTOS+TCP library.
To reproduce this result, we enabled the macro \texttt{ipconfigUSE\_NBNS} and
\texttt{ipconfigSUPPORT\_OUTGOING\_PINGS}, and ran fuzzing for 48 hours.
For the remaining four CVEs, we have studied their behaviors.
\texttt{CVE-2018-16598} is not crashable.
\texttt{CVE-2018-16526}, \texttt{CVE-2018-16528} and \texttt{CVE-2018-16522} 
can only be triggered with proper context
which we did not support in our taming.

\subsubsection{Comparison with Other Solutions}

We conducted experiments to compare our approach with related work.


\paragraph{Comparison with Emulation-based Solutions.}
We compared our approach with emulation-based solutions in general,
including a head-to-head comparison with HALucinator~\cite{ClementsGustafson2020halucinator}
on the same set of firmware samples.


We used a ready-to-use QEMU provided
by Zephyr that emulates the TI LM3S6965 platform~\cite{zephyrqemum3}.
Then, we applied the patch from TriforceAFL~\cite{TriforceAFL} to
this QEMU.
TriforceAFL is an AFL fuzzing tool with support for full-system emulation.
Finally we ran and fuzzed the MQTT library for Zephyr using
TriforceAFL.
The fuzzing lasted for 1 hour and on average the fuzzing throughput
was 23 test-cases per second.
This confirms the tremendous performance advantage of our approach over emulation-based
solutions. Similar performance measurement can be observed in relevant papers~\cite{P2IM,ClementsGustafson2020halucinator}.


\begin{table*}[t]
  \centering
    \caption{Comparison with HALucinator at a glance}
    \begin{adjustbox}{max width=\textwidth}
    \begin{tabular}{l|rrrrr|rrrrr}
    \toprule
    &  \multicolumn{5}{c|}{\textbf{HALunicator}} & \multicolumn{5}{c}{\textbf{Para-rehosting}} \\
    & Time & Executions & Paths & Crashes & Speed (\#/sec) & Time & Executions &  Paths & Crashes  & Speed (\#/sec) \\
    \midrule
    WYCINWYC  &  1d:00h &  1,548,582 &  612 &5 &17.92  & 11h:43m   & 27,326,874  & 3,166 & 909 & 647.86 \\
    Atmel lwIP HTTP (Ethernet) &  19d:04h & 37,948,954 & 8,081 &  273 & 22.92 & 12h:33m & 40,795,301  & 1,107  & 219 & 902.95\\
Atmel lwIP HTTP (TCP) &  0d:10h & 2,645,393 &  1,090 &  38 & 73.48 & 12h:00m  & 56,950,867 & 69 & 15 & 1,318.31\\
STM UDP Server &3d:08h &19,214,779 & 3,261 &0 & 66.72 &12h:00m & 38,979,912 & 621 & 16 & 902.31\\
STM UDP Client &3d:08h &12,703,448  &3,794 &0 & 44.11 &12h:00m & 53,785,098  & 599  & 65 & 1,245.03\\
STM TCP Server &3d:08h&16,356,129  &4,848 &0 & 56.79 &12h:00m   & 63,361,923  & 1,013   & 129   & 1,466.71\\
STM TCP Client  &3d:08h   &16,723,950  &5,012   &0 & 58.07 &12h:00m   & 47,192,271  & 1,222 & 58  & 1,092.41\\
STM ST-PLC     &1d:10h   &456,368 &772 &27 & 3.73 & 12h:15m & 112,579,017  & 323 & 41  & 2,552.81  \\                        
NXP TCP Server  &14d:00h  &218,214,107 &5,164   &0 & 180.40 &12h:02m & 38,316,493 & 448   & 0   &884.50\\
NXP UDP Server  &14d:00h  &240,720,229 &3,032   &0 & 199.01 &12h:00m & 36,186,349  & 264   &0   & 837.65\\
NXP HTTP Server &14d:00h  &186,839,871 &9,710   &0 & 154.46 &12h:39m & 65,724,013 & 1,101   & 0   & 1,443.22\\
    \bottomrule
    \end{tabular}%
    \end{adjustbox}
    
    \label{tab:HALucinatorCompare}%
\end{table*}%

We had a chance to compare HALucinator with ours head-to-head. We compiled and
tested the same set of firmware used in HALucinator from the GitHub
repo~\cite{HALucinatorsource}. However, the HALucinator authors did not
disclose the HAL versions in their experiments except the lwip version used in
Atmel firmware (1.4.1). For others, we used the latest SDK releases. We did
not test the two Atmel firmware images that use the 6LoWPAN interface, because we have not ported \sys to the 
Contiki OS. All the other samples are baremetal.
Moreover, since HALucinator targets binary firmware, it
feeds the test-cases in a non-standard way.
We had to tame the source code correspondingly to make sure
both treat the test-cases in exactly the same way.
Between each test-case, we used the AFL's built-in fork-server to
reset the firmware state.
Table~\ref{tab:HALucinatorCompare}
shows the results of the comparison, in which
the data of HALucinator were directly extracted from the original paper~\cite{ClementsGustafson2020halucinator}.
For all the samples, our approach overwhelms HALucinator 
in terms of execution speed, which is one of the most important factors in fuzzing.
We found that the
total paths found in HALucinator were generally higher than those found in \approach, even though we eventually triggered more crashes. This is because
HALucinator runs on top of QEMU. When a random external interrupt occurs, the
corresponding basic block transition is regarded as a new path. We acknowledge
that HALucinator minimizes this randomness by designing a deterministic timer.
That is, a timer interrupt is raised based on the number of executed basic blocks.
However, slightly different test-cases may cause a small change in execution path,
which further causes the timer to occur randomly. As a result, AFL may
mistakenly mark a path as new. This problem is avoided in our approach
because we model an MCU task as a POSIX thread, which
is oblivious of the emulated timer.

\begin{table}[t]
  \centering
    \caption{Accumulated coverage rate in comparison with HALucinator}
    \begin{adjustbox}{max width=\columnwidth}
    \begin{tabular}{l|rr}
    \toprule
    \multicolumn{1}{l|}{\textbf{Firmware}} & \textbf{Halucinator}  & \textbf{\Approach} \\
    \midrule
    WYCINWYC & 25.99\% & 27.91\% \\
    Atmel lwIP HTTP (Ethernet) & 47.65\% & 70.51\% \\
    Atmel lwIP HTTP (TCP) & 6.21\% & 6.26\% \\
    STM UDP Server & 28.37\% & 32.82\% \\
    STM UDP Client & 29.55\% & 38.22\% \\
    STM TCP Server & 40.31\% & 48.60\% \\
    STM TCP Client & 41.90\% & 56.88\% \\
    STM ST-PLC & 3.19\% & 25.05\% \\
    NXP TCP Server & 27.31\% & 40.66\% \\
    NXP UDP Server & 24.35\% & 24.96\% \\
    NXP HTTP Server & 45.83\% & 54.43\% \\
    \bottomrule
    \end{tabular}%
    \end{adjustbox}
    \label{tab:HALucinatorCompareRate}%
\end{table}%

In the above-mentioned experiment,
we compared the absolute values of certain important factors
such as the number of total paths.
However, due to the architecture differences,
these absolute numbers cannot sufficiently manifest 
the effectiveness of our solution.
To more convincingly illustrate the benefit of our approach,
we need to smooth the architecture differences.
To achieve this goal,
we re-ran fuzz testing on all the firmware images with both solutions
and collected more insightful details,
in particular, the accumulated basic block coverage \emph{rate}.
By using rates instead of absolute values, we ensure a fair comparison
on a best-effort basis.
Our experiments lasted for about 12 hours for each.
After fuzzing, we
replayed the generated test-case queues and
counted the accumulated basic block coverage rates.
When counting the basic block hit numbers,
we excluded the code for libc and HALs.
In other words, only the code for application and MCU libraries
is considered.
As shown in Table~\ref{tab:HALucinatorCompareRate},
we achieved higher basic block coverage rate in all the tested samples.
This is explained by the tremendous speedup brought by our approach.
Combined with the support of full-fledged ASAN,
our tool found more real bugs than HALucinator on the same set
of samples.
As an example,
for the STM ST-PLC firmware, we identified
nine bugs while HALucinator only identified one,
which is included in ours.
We have reported the eight new bugs to STMicroelectronics
and all of them have been confirmed and patched in the later release.

\paragraph{Comparison with On-device Analysis.}
No AFL-based fuzzing tool is available for real devices.
As a rough estimation, we only tested the overhead of fuzzing a real device.
We assume there is already some mechanisms to collect
path coverage information on the device for the AFL, \eg~ETM~\cite{etmv3}.


The overhead of fuzzing real devices is roughly composed of three parts,
test-case generation, test-case feeding, and program execution.
Here, we only estimated the overhead of test-case feeding and program execution, because
test-case generation should be the same for all the approaches.
We used the pyOCD~\cite{pyOCD} to control an NXP FRDM-K64F development board
and simulated the feeding process.
The python script writes a known test-case
into a reserved memory region of the board 1,000 times.
The program on the board terminates after the iteration is finished.
When transferring a test-case of 1 KB, it takes
about 0.23 seconds
for one test-case to be transferred.
We also compared the execution speed of the same workload
with \approach, Zephyr QEMU and the FRDM-K64F board. 
We found the execution speed of FRDM-K64F board
is 14 times slower than that of the QEMU
and 287 times slower than that of \approach.
In summary, the fuzzing speed on real devices 
is much lower than both our approach and emulation-based approaches.

\subsection{CVE Analysis}\label{bug_hunting}
We have
found 28 previously-unknown
bugs as shown in Table~\ref{new_bugs}.
Five of them were confirmed by CVE.
For the rest, we have reported them to the manufacturers or vendors
and 19 of them have been confirmed at the time of writing.
All the bugs were caused by memory errors and captured by ASAN.
Therefore, we argue that existing solutions without memory checker support, such
as emulation-based or on-device analysis, are less effective
in finding these bugs.

\begin{table}[t]
  \centering
    \caption{Previously-unknown bugs found}
    \begin{adjustbox}{max width=\columnwidth}
    \begin{tabular}{r|lllll}
    \toprule
    \multicolumn{1}{l|}{\textbf{ID}} & \textbf{RTOS}  & \textbf{Library} & \textbf{Bug type} & \textbf{Number} & \textbf{Status}\\
    \midrule
    \textbf{1} & FreeRTOS & MQTTv1 & Buffer Overflow & 1 & CVE-2019-13120  \\
    \textbf{2} & FreeRTOS & FATFS & Use After Free & 1 & CVE-2019-18178 \\
    \textbf{3} & FreeRTOS & uTasker Modbus & Buffer overflow & 5 & Confirmed \\
    \textbf{4} & FreeRTOS/Mbed OS & lwip & Buffer overflow & 3 & Confirmed \\
    \multirow{2}{*}{\textbf{5}} & \multirow{2}{*}{Mbed OS} & \multirow{2}{*}{MQTT} & Null Pointer & 1 & \multirow{2}{*}{CVE-2019-17210}\\
    &&&Dereference&\\
    \textbf{6} & Mbed OS & CoAP Parser & Buffer Overflow & 1 & CVE-2019-17212\\
    \textbf{7} & Mbed OS & CoAP Builder & Integer overflow & 1 & CVE-2019-17211 \\
    \textbf{8} & Mbed OS & Client-Cli & Buffer Overflow & 2 & Confirmed \\
    \textbf{9} & Mbed OS & Client-Cli & Off by one & 1 & Confirmed  \\
    \textbf{10} & LiteOS & MQTT & Buffer Overflow & 2 & submitted \\
    \textbf{11} & LiteOS & LWM2M & Use after free & 1 & submitted \\
    \textbf{12} & LiteOS & LWM2M & Buffer Overflow & 1 & submitted \\
    \textbf{13} & Bare-metal & STM-PLC & Buffer Overflow & 8 & Confirmed \\
    \bottomrule
    \end{tabular}%
    \end{adjustbox}
    \label{new_bugs}%
\end{table}%


\paragraph{\textbf{CVE-2019-13120}.}
The MQTT library lacks length checking for the received
publish message.
Attackers can manipulate this field so that
out-of-range memory access is triggered.
The authorized memory is later sent out through the acknowledge packet,
leading to a two-byte information leakage.

\paragraph{\textbf{CVE-2019-18178}.}
The file handler
freed by the function \texttt{ffconfigFREE()}
is reused by the function \texttt{FF\_FlushCache()}.
This bug can be immediately discovered using any test-case with the help of ASAN.
However, this library has never been executed against ASAN before and thus
the bug had not been identified for long.
This highlights the importance of the memory checker in
analyzing firmware.

\paragraph{\textbf{CVE-2019-17210}.}
The MQTT message can be manipulated so that an \texttt{if} statement
is invalided, leaving the default value of \texttt{mqttstring->lenstring.data} (\texttt{NULL})
unchanged. Later, this pointer is used.

\paragraph{\textbf{CVE-2019-17212}.}
The CoAP message is linearly parsed
by the function \texttt{sn\_coap\_parser\_options\_parse()} 
inside a while loop. 
However, inside the loop, the boundary of the message is not properly checked.

\paragraph{CVE-2019-17211.}
Two variables of type \texttt{uint16\_t} are added.
The result is used to allocate a buffer.
When the addition wraps around the maximum,
less memory is allocated than expected.

\paragraph{Others.}
Three bugs exist in the Client-Cli library of the Mbed OS.
This library parses the commands sent to the device.
If the command is manipulated, e.g., by removing the \texttt{NULL} character
or adding too many delimiters, buffer overflow or off-by-one bugs may occur.
The library uTasker Modbus has defined 
several global arrays with a length 
defined by the macro \texttt{MSG\_TIMER\_EVENT}. 
However, there is no offset check against 
the maximum length.
The library lwip copies the data from 
the data structure \textit{pbuf} which is 
implemented as a linked list.
It only copies from one sub payload 
with the length of the whole payload.
The MQTT library of LiteOS has an incorrect definition of an array
and does not check the validity of a pointer to the MQTT topic before accessing it.
The library LWM2M accesses a deallocated memory when closing the LWM2M session
and does not perform boundary check when traversing the received CoAP packet.
The STM-PLC application does not check the value of an increased or
decreased array index, leading to buffer overflows.

\input{tex/gap}

%% file: tex/gap.tex
\subsection{Identifying the Gap} 
\label{sec:gap}


Due to the architectural gap, we cannot 
guarantee that our results obtained on the rehosted program
can be reproducible on real MCU devices.
To identify this gap,
we developed a dataset comprised of vulnerable MCU OS libraries,
and tested different behaviors when running them
para-rehosted and on the real device.
We want to know if the bugs identified by our approach
are real bugs that can influence real devices.
During this experiment, we found that triggering a bug on the real device
rarely crashed the firmware immediately.
This observation agrees with previous work~\cite{what_you_corrupt}.
Therefore, we revisited this problem and further measured the different levels of
corruption observability on Linux machines and on real MCU devices.






\subsubsection{Dataset}

We crafted a corpus of code snippets with both 
the real bugs and artificial bugs. 
For real bugs, we included the nine re-producible bugs
in the FreeRTOS+TCP library originally reported by Zlab~\cite{zlabreport} (part 1),
plus the 28 new bugs discovered in this work (part 2).
For artificial bugs, we manually inserted eight types classic bugs in
an empty RTOS task (part 3).
Note that running code in part 3 should directly trigger the
bugs without inputting any test-case.
The dataset is listed in Table~\ref{tab:dataset}.




\subsubsection{Bug Re-producibility}
We manually fed the test-cases that can trigger bugs
on rehosted programs to the NXP FRDM-K64F board, and observed if the bug can be triggered.
We used the on-board CMSIS-DAP debug interface to track the execution and
manually verified that all the 28 bugs in part 2 of our dataset
can influence the real device.
This indicates that
the buggy source code can be easily manifested on any architecture.
However, the firmware did not crash immediately.
This is because the injected memory errors were not critical enough to
trigger observable hard faults. 
For these bugs, the firmware would crash non-deterministically in a long run, depending on the execution context.
This behavior imposes negative influences to bug hunting on real devices.
Specifically, during fuzzing, since the execution terminates immediately at exit points, these memory corruptions usually cannot cause observable crashes.

Regarding exploitability, due to the architecture differences (\eg~different
ISA and stack layout), a working proof of exploitation (PoE) on the rehosted
firmware might not work on real devices. Typically, PoE construction is
conducted case-by-case and bug exploitation is orthogonal to this work.


\subsubsection{Understanding the Bug Observability Gap}

\begin{table}[t]
  \centering
    \caption{Dataset and bug observability}
    \begin{adjustbox}{max width=\columnwidth}
    \begin{threeparttable}
    \begin{tabular}{r|lllrr}
    \toprule
    \multicolumn{1}{l|}{\multirow{2}[2]{*}{\textbf{ID}}} & \multirow{2}[2]{*}{\textbf{RTOS}} & \multirow{2}[2]{*}{\textbf{Library}} & \multirow{2}[2]{*}{\textbf{Bug type}} & \multicolumn{2}{c}{\textbf{Observability}} \\
          &       &       &       & \textbf{Para-rehosting} & \textbf{Real Board} \\
    \midrule
    \textbf{1} & FreeRTOS & FreeRTOS+TCP & Buffer Overflow & Y   & N \\
    \textbf{2} & FreeRTOS & FreeRTOS+TCP & Buffer Overflow & Y   & N \\
    \textbf{3} & FreeRTOS & FreeRTOS+TCP & Integer Underflow & Y   & N \\
    \textbf{4} & FreeRTOS & FreeRTOS+TCP & Buffer Overflow & Y   & N \\
    \textbf{5} & FreeRTOS & FreeRTOS+TCP & Buffer Overflow & Y   & N \\
    \textbf{6} & FreeRTOS & FreeRTOS+TCP & Buffer Overflow & Y   & N \\
    \textbf{7} & FreeRTOS & FreeRTOS+TCP & Buffer Overflow & Y   & N \\
    \textbf{8} & FreeRTOS & FreeRTOS+TCP & Buffer Overflow & Y   & N \\
    \textbf{9} & FreeRTOS & FreeRTOS+TCP & Buffer Overflow & Y   & N \\
    \midrule
    \textbf{10} & FreeRTOS & MQTT  & Buffer Overflow & Y   & N \\
    \textbf{11} & FreeRTOS & FATFS & Use After Free & Y   & N \\
    \textbf{12} & FreeRTOS & uTasker Modbus & Buffer Overflow & Y & N \\
    \textbf{13} & FreeRTOS & uTasker Modbus & Buffer Overflow & Y & N \\
    \textbf{14} & FreeRTOS & uTasker Modbus & Buffer Overflow & Y & N \\
    \textbf{15} & FreeRTOS & uTasker Modbus & Buffer Overflow & Y & N \\
    \textbf{16} & FreeRTOS & uTasker Modbus & Buffer Overflow & Y & N \\
    \textbf{17} & FreeRTOS & lwip & Buffer Overflow & Y & N \\
    \textbf{18} & FreeRTOS & lwip & Buffer Overflow & Y & N \\
    \textbf{19} & FreeRTOS & lwip & Buffer Overflow & Y & N \\
    \textbf{20} & MbedOS & MQTT  & Null Pointer & Y   & N \\
    \textbf{21} & MbedOS & CoAP Parser & Buffer Overflow & Y   & N \\
    \textbf{22} & MbedOS & CoAP Builder & Integer Overflow & Y   & N \\
    \textbf{23} & MbedOS & Client-Cli & Buffer Overflow & Y   & N \\
    \textbf{24} & MbedOS & Client-Cli & Buffer Overflow & Y   & N \\
    \textbf{25} & MbedOS & Client-Cli & Off By One & Y   & N \\
    \textbf{26} & LiteOS & MQTT & Buffer Overflow & Y & N \\
    \textbf{27} & LiteOS & MQTT & Buffer Overflow & Y & N \\
    \textbf{28} & LiteOS & LWM2M & Use after free & Y & N \\
    \textbf{29} & LiteOS & LWM2M & Buffer Overflow & Y & N \\
    \textbf{30} & Baremetal & STM-PLC & Buffer Overflow & Y   & N \\
    \textbf{31} & Baremetal & STM-PLC & Buffer Overflow & Y   & N \\
    \textbf{32} & Baremetal & STM-PLC & Buffer Overflow & Y   & N \\
    \textbf{33} & Baremetal & STM-PLC & Buffer Overflow & Y   & N \\
    \textbf{34} & Baremetal & STM-PLC & Buffer Overflow & Y   & N \\
    \textbf{35} & Baremetal & STM-PLC & Buffer Overflow & Y   & N \\
    \textbf{36} & Baremetal & STM-PLC & Buffer Overflow & Y   & N \\
    \textbf{37} & Baremetal & STM-PLC & Buffer Overflow & Y   & N \\
    \midrule
    \textbf{38} & FreeRTOS & n/a     & Div By Zero & Y   & Config \\
    \textbf{39} & MbedOS & n/a     & Div By Zero & Y   & Config \\
    \textbf{40} & FreeRTOS & n/a     & Integer Overflow & Late  & N \\
    \textbf{41} & MbedOS & n/a     & Integer Overflow & Late  & N \\
    \textbf{42} & FreeRTOS & n/a     & Stack Overflow & Y   & N \\
    \textbf{43} & MbedOS & n/a     & Stack Overflow & Y   & N \\
    \textbf{44} & FreeRTOS & n/a     & Heap Overflow & Partially   & Partially \\
    \textbf{45} & MbedOS & n/a     & Heap Overflow & Y   & N \\
    \textbf{46} & FreeRTOS & n/a     & NULL Pointer & Y   & Partially \\
    \textbf{47} & MbedOS & n/a     & NULL Pointer & Y   & Partially \\
    \textbf{48} & FreeRTOS & n/a     & Double Free & Y   & heap4 only \\
    \textbf{49} & MbedOS & n/a     & Double Free & Y   & Y \\
    \textbf{50} & FreeRTOS & n/a     & Use After Free & heap3 only & N \\
    \textbf{51} & MbedOS & n/a     & Use After Free & Y   & N \\
    \textbf{52} & FreeRTOS & n/a     & Format String & N    & N \\
    \textbf{53} & MbedOS & n/a     & Format String  & N    & N \\
    \bottomrule
    \end{tabular}%
    \textbf{Config} means certain hardware feature needs to be enabled to observe the crash.
    \textbf{Late} means the crash is not immediately observable but can be observed later.
    \textbf{Partially} means the observability depends on the concrete context.
    \textbf{heap4} and \textbf{heap3} are FreeRTOS specific.
    They are two different heap implementations. 
    \end{threeparttable}
    \end{adjustbox}
    \label{tab:dataset}%
\end{table}%

The bug observability issue on embedded systems was firstly explained
by Muench et. al.~\cite{what_you_corrupt}.
It imposes a huge challenge in fuzzing firmware on real devices.
As mentioned before, we observed the same phenomenon.
In this section, we try to conduct experiments to
better understand the bug observability gap on Linux machines and real MCU devices.

Our approach is to feed the same test-cases that trigger bugs on the rehosted programs to the real device.
This only applies for part 1 and 2 of the dataset because part 3 does not need
any input.
For bug ID 11 and 28, any test-case can trigger the UAF vulnerability.
As shown in Table~\ref{tab:dataset},
most bugs can be observed immediately on the rehosted programs thanks to the ASAN support while
most bugs cannot be observed on real devices.
In that follows, we explain some interesting findings.

\paragraph{Div by Zero.}
By default, diving by zero yields zero on the FRDM-K64F board.
As a result, the board will continue execution
without  crashing.
We later found that ARM processor needs to explicitly enable
div by zero detection by setting the \textit{Configuration and Control Registers} (CCR).
After the register was set properly,
we could observe this bug instantly.


\paragraph{Stack Overflow.}
On the rehosted programs, we could observe stack overflow 
with the help of ASAN easily.
On the real board, with the default configuration,
the firmware will continue execution.
However, this problem can be mitigated by configuring stack overflow detection.

For FreeRTOS, it provides a macro called \texttt{configCHECK\_FOR\_STACK\_OVERFLOW}
to enable stack overflow detection~\cite{zlabreport}.
When configured to 1, the FreeRTOS kernel will
check if the processor stack pointer remains within the valid stack space
when the task is swapped out.
When configured to 2, the FreeRTOS kernel will
initialize known values at the end of the stack.
When the task is swapped out, the kernel checks
whether the known values have been corrupted.
Obviously, both mechanisms cannot guarantee to catch all stack overflows,
and may delay the detection until the task is swapped out.





The Mbed OS provides stack overflow detection
via stack stats tracking~\cite{mbedos_mem_tracing}.
When enabled, stack overflow can be observed by monitoring stack usage.

\paragraph{Null Pointer Dereference.}
On rehosted programs, we could observe null pointer dereferences easily because
accessing memory at zero always causes a segment fault -- the OS maps the corresponding virtual pages as unpresent.
On a real board, in particular the ARM Cortex-m MCU devices,
read and write to \texttt{NULL} address lead to different results.
This is because address zero is typically where the ROM resides.
Therefore, reading from it is allowed
while writing to it causes an escalated hard fault.
For example, when we fed the test-case triggering \texttt{CVE-2019-17210} to the real device,
the real device never crashed. However, the firmware execution had been in a false state.

%% file: tex/6-Disc.tex
\section{Comparison with Other Work}

\begin{figure*}[t]
\centering
\includegraphics[width=.8\textwidth]{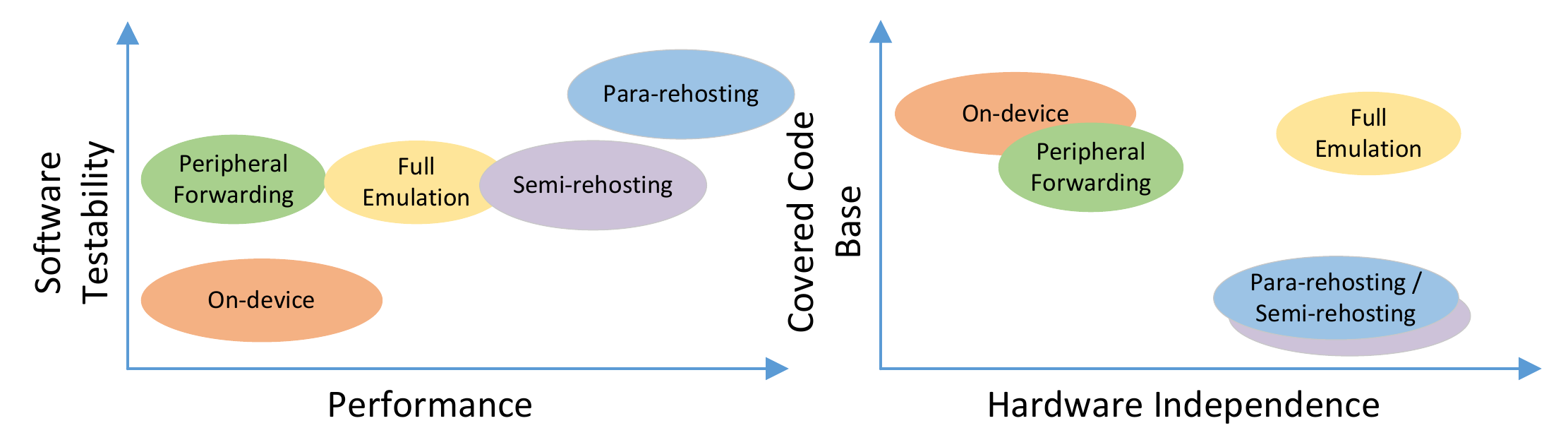}
\caption{Comparison of different approaches in terms of performance,
software testability, hardware independence and covered code base.}
\label{fig:comparision}
\end{figure*}


\label{sec:comparision}
To examine the security of MCU firmware, several approaches
have been discussed.
We categorize them into four classes.

\begin{enumerate}
	\item \textbf{On-device Analysis:}
	This approach requires real devices to do the testing.
	It gets the most authentic results, but is
	less scalable and lack visibility.
	It is hard to collect execution information on a bare-metal machine.
	
	\item \textbf{Full Emulation:}
	To overcome the performance and scalability issues on real hardware,
	researchers proposed using a full emulator, such as QEMU, 
	to emulate firmware execution.
	The main challenge is to emulate different
	peripherals and there has been some
	efforts towards this goal~\cite{P2IM,PRETENDER}.
	Theoretically, this approach 
	achieves better visibility
	to the firmware execution. 
    Unfortunately, as far as we know, no existing work is able to precisely
    support previously-unknown devices.
    Therefore, at the current state, this approach only exists in an idealistic setting.
	
	\item \textbf{Peripheral Forwarding:}
	As a middle ground solution, a hybrid approach forwards peripheral
	accesses to real devices and runs the firmware inside an
	emulator~\cite{AVATAR,what_you_corrupt,SURROGATES}.
	However, performance and scalability issues
	are still unsolved due to the dependence on real hardware.

	\item \textbf{Semi-rehosting:}  
	In semi-rehosting solutions like Halucinator~\cite{ClementsGustafson2020halucinator}, the main logic of
	the firmware is still executed inside an emulator. However, high-level HAL
	functions are identified and replaced with rehosted handlers on the host. As
	such, complex modeling of diverse peripherals is avoided.

\end{enumerate}

\paragraph{Comparison.} In Figure~\ref{fig:comparision}, we compare our approach (\approach)
with existing work from different dimensions.
Note that
a perfect full emulation solution that can precisely emulate
arbitrary MCU devices \emph{does not} exist so far.

In terms of performance, even high-end MCUs run
at a lower frequency. Therefore, on-device testing could be less efficient.
Moreover, parallelism is limited on real devices. To add a system under test (SUT),
a new real device must be integrated and be coordinated with others properly.
As for peripheral forwarding, frequent rebooting of the device 
and signal forwarding are both time-consuming.
Emulation-based approaches (including semi-rehosting) can run multiple SUTs simultaneously on the host,
but instruction translation incurs unavoidable overhead. 
Our approach enables native testing on the host, fully benefiting
from the parallelism and high performance of modern hardware.

Regarding software testability, we mean the capability to
observe the behavior of an SUT. For example, being able to
observe a crash is vital for fuzzing.
As another example, collecting the run-time execution path
can aid fuzzing by generating high-quality test-cases.
To improve testability, instrumentation is often used.
Instrumentation allows for collecting more context-rich information.
In fuzzing, instrumentation tools like ASAN~\cite{asan}
can improve the crash identification capability significantly.
On a real device, it is nearly impossible to collect
firmware execution information, unless non-intrusive hardware debugging features,
such as ETM are in-place~\cite{etmv3}.
Peripheral forwarding, full emulation, and semi-rehosting all rely on 
an emulator such as QEMU. With an indirection layer, testability can
be greatly improved.
However, it is still incomparable with native rehosting.
Specifically, in practice, emulation-based approaches can only
extract information in the context of the emulated machine,
whereas native rehosting enables information collection
in the context of the program itself, e.g., by instrumentation.
In our evaluation (section~\ref{bug_hunting}), we have demonstrated
how ASAN could improve the bug finding capability.
For all the bugs we identified using our tool except one (CVE-2019-17210),
if ASAN was not enabled, the bugs became invisible.

In terms of hardware-independence,
both on-device analysis and peripheral forwarding require
a real device, whereas
\approach and emulation-based approaches
do not depend on any real device.
As such, they are more scalable.

As for code coverage, all the approaches could
cover hardware-independent code, which is enclosed in red lines in Figure~\ref{fig:stack}.
However,
because \approach and semi-rehosting replace
hardware-dependent code with native implementations, they cannot find problems with
low-level hardware related code (\eg~drivers), which
is enclosed in green lines in Figure~\ref{fig:stack}.
Full emulation has the potential to cover the whole software stack.
However, {P$^{2}$IM}~\cite{P2IM}, a representative full emulation approach,
only provides approximate emulation, which does not suffice for discovering driver bugs.
On the other hand, peripheral forwarding
and on-device analysis can capture hardware behaviors with high fidelity 
and therefore can support testing diver code.

Last but not least, due to the nature of cross-compilation and the need for
instrumentation, our approach requires the source code, while
others can work directly on the binary.
As such, \approach is most suitable for chip vendors or third-party MCU OS vendors to do in-house testing, rather than for hackers or researchers to find bugs in binaries. 
Note that even if these vendors have access to the source code, they often lack an efficient dynamic testing tool in their arsenal (\eg~ASAN for MCU code). Our tool bridges this gap.


%% file: tex/8-avl.tex


\section{Related Work}
\subsection{MCU OS Simulator}\label{sec:nativecomparison}
To create a smooth and pleasant experience for MCU developers,
some MCU OSs provide a simulator that enables developers to write,
run, and debug code without acquiring real hardware.
Typically, there are three methods adopted.

\subsubsection{Web-based Simulator}
Mbed OS 5 provides a cross-compile simulator based on
Emscripten~\cite{emscripten} and Node.js. 
In essence, C/C++ code is translated to
WebAssembly that can run on browsers. 
This approach provides the most convenience since the developers
do not need to install the simulator environment. 
However, it is incompatible with existing dynamic testing tools.

\subsubsection{QEMU-based Emulator}
Typically, QEMU-based emulators
only emulate a specific board.
Supporting arbitrary boards requires more engineering efforts,
which is unmanageable given the huge amount of MCU chips.
As a result, to allow the developers to fully test the APIs of the RTOS, 
the provided emulation board should emulate a wide range of supported peripherals,
which is not the case in practice.
For example, we found that only a tiny portion of OS APIs can be tested
on the RT-Thread emulator~\cite{rtthreadqemu}.
QEMU-based approach is compatible with (patched) AFL~\cite{TriforceAFL} with reduced speed.
But ASAN cannot be fully supported.
This approach is adopted by RT-Thread~\cite{rtthreadqemu} and  Zephyr~\cite{zephyrqemu}.


\subsubsection{Native Simulator}
Sometimes, the MCU OS developers provide a simulator that
when compiled with the application and the kernel,
can generate  a native application on the host OS.
This kind of simulator typically accommodates specific
characteristics of the target MCU OS and
has certain optimizations to it.
The introduced optimization may influence the intended
behavior of the firmware.
For example, to bring a deterministic
environment for developing and debugging,
the native simulator provided by Zephyr~\cite{zephyremu}
models native execution to be
infinitely fast. As a result, interrupts, including timers, cannot arbitrarily
suspend a thread. The firmware has to explicitly unmask them if they were
pending. This indicates this modeling cannot faithfully simulate the complex
interactions between the firmware and hardware. 
Race condition issues caused by concurrency and scheduling cannot be discovered faithfully by this approach. 
On the contrary,
our work faithfully simulates the non-deterministic nature
of the hardware and allows interrupts to kick in at any time.
Therefore, it can capture various real-world software issues. 

Ad-hoc simulator development does not consider generalization
and the developed simulator cannot work for another MCU OS.
For example, we found that the Zephyr simulator 
has a hardware model adaptation layer that is deeply
coupled with the Zephyr drivers and thus cannot be easily
re-used to simulate other MCU OSs.
Our work abstracts common behaviors of the hardware
and correspondingly provides unified backends to
simulate the firmware execution, minimizing the needed
re-engineering effort to support other OSs.
In addition, \approach not only enables the simulation of the MCUs,
but also peripherals.
Finally, \approach shares two key benefits with other native simulators:
1) improved execution speed for efficient testing,
and 2) better instrumentation capability to disclose program issues.
Native simulator is supported by NuttX~\cite{nuttx}, FreeRTOS~\cite{freertoswin}, Zephyr~\cite{zephyremu}, etc.

\subsection{Generic Firmware Emulation}
In {P$^{2}$IM}~\cite{P2IM}, the authors propose to abstract
a model for a class of MCUs based on device datasheets or processor documentation.
Then {P$^{2}$IM} instantiates 
the abstract model automatically with the firmware-specific information.
As such, {P$^{2}$IM} is oblivious to peripheral designs and generic to
firmware implementations.
Moreover, it channels inputs from the AFL to continuously fuzz
the emulated firmware.
PRETENDER~\cite{PRETENDER} ``learns''
the interaction model between the original hardware and the firmware,
and automatically creates models of peripherals.
Laelaps~\cite{laelaps} addresses a similar problem using symbolic execution.
HALucinator~\cite{ClementsGustafson2020halucinator} avoids the problem of peripheral 
emulation by replacing the high-level HAL function with a host implementation.
All these solutions rely QEMU for architectural emulation and therefore
suffer from lower performance and bad testability.
Since they directly test the binary-form firmware, no source code is needed.

Conceptually, both HALucinator and \approach replace HAL functions with native
implementations.  However, the motivation and method are quite different. 
HALucinator directly matches and hooks HAL functions in binaries. It is
helpful for third-party researchers to find bugs in binaries. To build the
function matching database, it needs the HAL source code to calculate the
signatures of the underlying HAL functions before-hand. The matching results
are subject to inaccuracy caused by collision, missing functions, etc.
\Approach inherently needs the source code and relies on some human efforts to
craft glue layers. However, we alleviate this problem by providing
ready-to-use glue layers for popular HALs. For others, common para-APIs and
templates are provided.  In evaluation, we have demonstrated the tremendous
performance advantage to HALucinator brought by native rehosting. Moreover, we
found that all the tested sample in HALucinator are baremetal. It remains
unknown whether HALucinator can handle more complex OS libraries. For example,
the authors state that to support RIOT OS, they had to manually implement the
context switching as a handler. To facilitate fuzzing, they also need to write
additional modules to accommodate testcase input, monitor execution results,
increase crash visibility (\eg~an ASAN-fashioned heap memory tracker), etc.


\subsection{Peripheral Forwarding}
Researchers also propose a hybrid emulation approach in which the real hardware
is used to handle peripheral operations.
Avatar~\cite{AVATAR} and Avatar2~\cite{avatar2} propose a dynamic analysis framework that
executes the firmware in QEMU until an I/O request is encountered.
In this case, the request is forwarded to the real hardware.
PROSPECT~\cite{kammerstetter2014prospect} introduces a novel approach that involves the 
partial emulation of an embedded device's firmware during a fuzzing experiment.
By forwarding system calls that are likely to access peripherals,
this approach can emulate Linux-based embedded systems.
SURROGATES~\cite{SURROGATES} improves Avatar 
by using a custom, low-latency FPGA bridge between the host’s PCI Express bus
and the system under test.
It also allows the emulator full access to the system’s peripherals.
These approaches still rely on real hardware and thus is not scalable.
The performance cannot exceed those achieved by QEMU- or rehosting-based solutions.


\subsection{Firmware Analysis}
Symbolic execution is commonly used in analyzing MCU firmware.
FIE~\cite{davidson2013fie} leverages the KLEE~\cite{cadar2008klee} symbolic execution engine 
to provide an extensible platform 
for detecting firmware bugs. 
FIE does not simulate hardware interaction.
That is, writes to a peripheral are ignored 
and reads return unconstrained symbolic values. 
Moreover, it is specific to the MSP430 family microcontrollers.
FirmUSB~\cite{FirmUSB} analyzes embedded USB devices and uses domain knowledge
to speed up the symbolic execution of firmware.
Compared to unconstrained symbolic execution, FirmUSB can improve the
performance by a factor of seven.
Inception~\cite{corteggiani2018inception} is another KLEE-based
system aiming at testing a complete firmware image.
It symbolically executes LLVM-IR merged from source code, assembly, and binary libraries. 
To handle peripherals, it either follows the approach of FIE or redirects the read operation to
a real device. 
Both \approach and Inception need the source code.
Although we only used fuzzing to find bugs in this work, there
is no technical obstacle of using symbolic execution tools
such as angr~\cite{angr} to analyze 
the rehosted program or even use hybrid fuzzing~\cite{217563,stephens2016driller} to improve
efficiency.

Previous work has made tremendous progress in analyzing Linux-based
firmware~\cite{Firmadyne,firmafl}. The high-level idea is to design a
generic kernel for all the devices. This approach leverages the abstract layer
offered by the Linux kernel, but cannot work for the MCU firmware where the
kernel and tasks are mingled together. Finally, for PLCs, Almgren et al.
developed several mutation-based and generational-based fuzzers against
various PLCs and smart meters~\cite{almgrend5}. They discovered several known
and unknown denial of service vulnerabilities.

\subsection{OS Customization}

The need for better performance and security has pushed OS customization
techniques in recent years~\cite{exokernel,unikernel,IX,arrakis}. For example,
Exokernel~\cite{exokernel} provides a minimal set of hardware-level interfaces
for multiplexing hardware resources among applications. On top of it, each
application implements a library OS (libOS) that include a customized and
optimized OS abstraction. Similarly, Unikernel~\cite{unikernel} compiles a
highly specialized libOS with the application, removing unnecessary functions
in commodity OSs. 
The approach used in
\approach is aligned  with this new trend in OS design. Specifically, we also
abstract a common and minimal set of hardware interfaces for MCU OSs. Each MCU
OS implements its designed functionality based on
this common hardware interface in different ways.



%% file: tex/7-con.tex
\section{Conclusions}

In-house security testing of MCU firmware is crucial for IoT security.
However, due to the different testing environment,
sophisticated tools in x86 are not available for MCU firmware.
Re-compiling the firmware to the native x86 host can directly
address this problem.
However, ad-hoc porting is daunting, prone to errors
and sometimes impossible. 
We therefore propose \approach to ease this
process. The portable MCU is able to model the common functions of an MCU
while para-APIs facilitate HAL-based peripheral function replacement to deal
with peripherals. Rehosting MCU OSs directly addresses fundamental issues
(performance, scalability and visibility) faced by  existing solutions. 
We have implemented our idea and rehosted nine OSs for MCU.
We did security testing for libraries of Amazon FreeRTOS, ARM Mbed OS,
Zephyr and LiteOS. Most libraries shipped with these OSs can be tested by
off-the-shelf dynamic analysis tools, including AFL and ASAN. 
Our experiments suggested that the bugs in the real firmware
are re-producible in rehosted firmware.
And the bugs are more observable on rehosted firmware.
Running our tool with
fuzzing, previously-unknown bugs have been discovered.

%% file: tex/app.tex
\clearpage
\label{sec:libsupport}
\begin{sidewaystable}[t]
  \centering
    \caption{Library support}
    \begin{adjustbox}{max width=\textwidth}
    \begin{tabular}{l|p{.45\textwidth}|p{.3\textwidth}}
    \toprule
    \multicolumn{1}{l|}{\textbf{RTOS}} & \multicolumn{1}{c|}{\textbf{Supported Libraries}} & \multicolumn{1}{c}{\textbf{Unsupported Libraries}} \\
    \midrule
    \textbf{FreeRTOS} & TCP/IP, atomic operations, linear containers, logging, static memory, task pool, device defender, device shadow, greengrass, MQTTv2.0.0, MQTTv1.0.0, HTTPS, PKCS\#11, secure sockets, TLS, FATFS & bluetooth low energy, over-the-air (OTA) agent, WiFi \\
    \midrule
    \textbf{Mbed OS} & mbed-crypto, mbedtls, mbed-edge, littlefs, tinycbor, mbed-trace, mbed-cloud-client, mbed-coap, mbed-client-cli, cose-c, lwip, sal, mbed-spiffs & mbed-bootloader, sal-stack-nanostack-eventloop, Arm Mbed uVisor, swo \\
    \midrule
    \textbf{Zephyr} & file systems, logging, BSD sockets, DNS resolve, IPv4/IPv6 primitives and helpers, network management, network statistics, network timeout, SNTP, SOCKS5 proxy, trickle timer library, websocket client, CoAP, LWM2M, MQTT, gPTP, PTP time format & bluetooth, display interface \\
    \midrule
    \textbf{LiteOS} & MQTT, LWM2M, nb-iot, fs, cJSON, lwip, mbedtls & nnacl, gui, OTA, sensorhub \\
    \bottomrule
    \end{tabular}%
    \end{adjustbox}
  \label{tab:libsupport_tab}%
\end{sidewaystable}%

\begin{sidewaystable}[t]
  \centering
  \begin{threeparttable}
    \caption{Porting details for selected RTOSs}
    \begin{tabular}{c|l|lll|r}
    \toprule
          & \textbf{Module} & \multicolumn{1}{l}{\textbf{Upstream}} & \textbf{Downstream} & \textbf{Common Backend} & \textbf{Glue Layer (LOC)}\\
    \midrule
    \multirow{4}[2]{*}{\textbf{FreeRTOS}} & Task Creation &       & pxPortInitialiseStack & PMCU\_BE\_Task\_Create & 5\\
          & SysTick & \multicolumn{1}{l}{vTaskSwitchContext} &  n/a   & PMCU\_BE\_Systick\_Handler & 12\\
          & \multirow{2}[1]{*}{Synchronization} &       & portDISABLE\_INTERRUPTS & PMCU\_BE\_Disable\_Irq & 3\\
          &       &       & portENABLE\_INTERRUPTS & PMCU\_BE\_Enable\_Irq & 3\\
    \midrule
    \multirow{4}[2]{*}{\textbf{Mbed OS}} & Task Creation &       & svcRtxThreadNew & PMCU\_BE\_Task\_Create & 5\\
          & SysTick & \multicolumn{1}{l}{osRtxTick\_Handler} &  n/a   & PMCU\_BE\_Systick\_Handler & 22\\
          & \multirow{2}[1]{*}{Synchronization} &       & \_\_disable\_irq & PMCU\_BE\_Disable\_Irq & 3\\
          &       &       & \_\_enable\_irq & PMCU\_BE\_Enable\_Irq & 3\\
    \midrule
    \multirow{4}[2]{*}{\textbf{Zephyr}} & Task Creation &       & z\_arch\_new\_thread & PMCU\_BE\_Task\_Create & 10\\
          & SysTick & \multicolumn{1}{l}{z\_reschedule} &  n/a   & PMCU\_BE\_Systick\_Handler & 25\\
          & \multirow{2}[1]{*}{Synchronization} &       & z\_arch\_irq\_disable & PMCU\_BE\_Disable\_Irq & 3\\
          &       &       & z\_arch\_irq\_enable & PMCU\_BE\_Enable\_Irq & 3\\
    \midrule
    \multirow{4}[2]{*}{\textbf{LiteOS}} & Task Creation &       & OsTaskStackInit & PMCU\_BE\_Task\_Create & 13\\
          & SysTick & \multicolumn{1}{l}{OsTaskSchedule} &  n/a  & PMCU\_BE\_Systick\_Handler & 14\\
          & \multirow{2}[1]{*}{Synchronization} &       & LOS\_HwiDisable & PMCU\_BE\_Disable\_Irq & 3\\
          &       &       & LOS\_HwiEnable & PMCU\_BE\_Enable\_Irq & 3\\
    \bottomrule
    \end{tabular}%
    \begin{tablenotes}
    \item The SysTick module does not have any downstream
    function because the hooks have been placed during task creation. The upstream functions are used for task scheduling.
    \end{tablenotes}
  \label{case_study}%
  \end{threeparttable}
\end{sidewaystable}%

\clearpage
\onecolumn

{
\fontsize{5.4}{5.4}\selectfont
\begin{longtable}{p{5em}|l|p{16em}p{16em}p{16em}}
\caption{Supported peripherals and the corresponding frontend/backend functions}\\
\toprule
\textbf{Category} & \textbf{Peripheral} & \textbf{STM32 HAL (frontend)} & \textbf{NXP HAL (frontend)} & \textbf{Para-API (backend)}  \\
\midrule
\endhead
\midrule
\multicolumn{5}{r}{Continued on next page} \\
\endfoot
\bottomrule
\endlastfoot
\textbf{I/O} &{UART} & HAL\_UART\_Init() &{UART\_Init()} &{HAL\_BE\_IO\_return\_success()} \\
          &       & HAL\_HalfDuplex\_Init() &{UART\_Deinit()} &  \\
          &       & HAL\_LIN\_Init() &       &  \\
          &       & HAL\_MultiProcessor\_Init() &       &  \\
          &       & HAL\_RS485Ex\_Init() &       &  \\
          &       & HAL\_UART\_DeInit() &       &  \\
\cmidrule{3-5}          &       & HAL\_UART\_Receive\_DMA() &{UART\_ReadBlocking()} &{HAL\_BE\_IO\_read()} \\
          &       & HAL\_UART\_Receive\_IT() &{UART\_ReadNonBlocking()} &  \\
          &       & HAL\_UART\_Receive() &{UART\_TransferReceiveNonBlocking()} &  \\
\cmidrule{3-5}          &       & HAL\_UART\_Transmit\_DMA() &{UART\_WriteBlocking()} &{HAL\_BE\_IO\_write()} \\
          &       & HAL\_UART\_Transmit\_IT() &{UART\_WriteNonBlocking()} &  \\
          &       & HAL\_UART\_Transmit() &{UART\_TransferSendNonBlocking()} &  \\
\cmidrule{2-5}          &{I2C} & HAL\_I2C\_Init() &{I2C\_MasterInit()} &{HAL\_BE\_IO\_return\_success()} \\
          &       & HAL\_I2C\_DeInit() &{I2C\_MasterDeinit()} &  \\
          &       &       &{I2C\_SlaveInit()} &  \\
          &       &       &{I2C\_SlaveDeinit()} &  \\
\cmidrule{3-5}          &       & HAL\_I2C\_Master\_Receive\_DMA() &{I2C\_MasterReadBlocking()} &{HAL\_BE\_IO\_read()} \\
          &       & HAL\_I2C\_Master\_Receive\_IT() &{I2C\_SlaveReadBlocking()} &  \\
          &       & HAL\_I2C\_Master\_Receive() &       &  \\
          &       & HAL\_I2C\_Master\_Seq\_Receive\_DMA() &       &  \\
          &       & HAL\_I2C\_Master\_Seq\_Receive\_IT() &       &  \\
          &       & HAL\_I2C\_Mem\_Read\_DMA() &       &  \\
          &       & HAL\_I2C\_Mem\_Read\_IT() &       &  \\
          &       & HAL\_I2C\_Mem\_Read() &       &  \\
          &       & HAL\_I2C\_Slave\_Receive\_DMA() &       &  \\
          &       & HAL\_I2C\_Slave\_Receive\_IT() &       &  \\
          &       & HAL\_I2C\_Slave\_Receive() &       &  \\
          &       & HAL\_I2C\_Slave\_Seq\_Receive\_DMA() &       &  \\
          &       & HAL\_I2C\_Slave\_Seq\_Receive\_IT() &       &  \\
\cmidrule{3-5}          &       & HAL\_I2C\_Master\_Transmit\_DMA() &{I2C\_MasterWriteBlocking()} &{HAL\_BE\_IO\_write()} \\
          &       & HAL\_I2C\_Master\_Transmit\_IT() &{I2C\_SlaveWriteBlocking()} &  \\
          &       & HAL\_I2C\_Master\_Transmit() &       &  \\
          &       & HAL\_I2C\_Master\_Seq\_Transmit\_DMA() &       &  \\
          &       & HAL\_I2C\_Master\_Seq\_Transmit\_IT() &       &  \\
          &       & HAL\_I2C\_Mem\_Write\_DMA() &       &  \\
          &       & HAL\_I2C\_Mem\_Write\_IT() &       &  \\
          &       & HAL\_I2C\_Mem\_Write() &       &  \\
          &       & HAL\_I2C\_Slave\_Transmit\_DMA() &       &  \\
          &       & HAL\_I2C\_Slave\_Transmit\_IT() &       &  \\
          &       & HAL\_I2C\_Slave\_Transmit() &       &  \\
          &       & HAL\_I2C\_Slave\_Seq\_Transmit\_DMA() &       &  \\
          &       & HAL\_I2C\_Slave\_Seq\_Transmit\_IT() &       &  \\
\cmidrule{2-5}          &{SPI} & HAL\_SPI\_Init() &{DSPI\_MasterInit()} &{HAL\_BE\_IO\_return\_success()} \\
          &       & HAL\_SPI\_DeInit() &{DSPI\_SlaveInit()} &  \\
          &       &       &{DSPI\_Deinit()} &  \\
\cmidrule{3-5}          &       & HAL\_SPI\_Receive\_DMA() &{DSPI\_ReadData()} &{HAL\_BE\_IO\_read()} \\
          &       & HAL\_SPI\_Receive\_IT() &       &  \\
          &       & HAL\_SPI\_Receive() &       &  \\
          &       & HAL\_SPI\_TransmitReceive\_DMA() &       &  \\
          &       & HAL\_SPI\_TransmitReceive\_IT() &       &  \\
          &       & HAL\_SPI\_TransmitReceive() &       &  \\
\cmidrule{3-5}          &       & HAL\_SPI\_Transmit\_DMA() &{DSPI\_MasterHalfDuplexTransferBlocking()} &{HAL\_BE\_IO\_write()} \\
          &       & HAL\_SPI\_Transmit\_IT() &{DSPI\_MasterHalfDuplexTransferNonBlocking()} &  \\
          &       & HAL\_SPI\_Transmit() &{DSPI\_MasterTransferBlocking()} &  \\
          &       & HAL\_SPI\_TransmitReceive\_DMA() &{DSPI\_MasterTransferNonBlocking()} &  \\
          &       & HAL\_SPI\_TransmitReceive\_IT() &{DSPI\_MasterWriteData()} &  \\
          &       & HAL\_SPI\_TransmitReceive() &{DSPI\_MasterWriteDataBlocking()} &  \\
          &       &       &{DSPI\_SlaveTransferNonBlocking()} &  \\
          &       &       &{DSPI\_SlaveWriteData()} &  \\
          &       &       &{DSPI\_SlaveWriteDataBlocking()} &  \\
\cmidrule{2-5}          &{Ethernet} & HAL\_ETH\_Init() &{ENET\_Init()} &{HAL\_BE\_NetworkInit()} \\
          &       & HAL\_ETH\_DeInit() &{ENET\_Deinit()} &  \\
\cmidrule{3-5}          &       & HAL\_ETH\_GetReceivedFrame\_IT() &{ENET\_ReadFrame()} &{HAL\_BE\_NetworkReceive()} \\
          &       & HAL\_ETH\_GetReceivedFrame() &{ENET\_ReadFrameMultiRing()} &  \\
\cmidrule{3-5}          &       & HAL\_ETH\_TransmitFrame() &{ENET\_SendFrame()} &{HAL\_BE\_NetworkSend()} \\
          &       &       &{ENET\_SendFrameMultiRing()} &  \\
\cmidrule{2-5} &{ADC} & HAL\_ADC\_Init() &{ADC16\_Init()} &{HAL\_BE\_IO\_return\_success()} \\
          &       & HAL\_ADC\_DeInit() &{ADC16\_Deinit()} &  \\
\cmidrule{3-5}          &       & HAL\_ADC\_GetValue() &{ADC16\_GetChannelConversionValue()} &{HAL\_BE\_IO\_read()} \\
          &       & HAL\_ADC\_Start\_DMA() &       &  \\
          &       & HAL\_ADC\_Start\_IT() &       &  \\
          &       & HAL\_ADC\_Start() &       &  \\
\cmidrule{2-5}          &{SAI} & HAL\_SAI\_Init() &{SAI\_Init()} &{HAL\_BE\_IO\_return\_success()} \\
          &       & HAL\_SAI\_DeInit() &{SAI\_Deinit()} &  \\
\cmidrule{3-5}          &       & HAL\_SAI\_Receive\_DMA() &{SAI\_ReadBlocking()} &{HAL\_BE\_IO\_read()} \\
          &       & HAL\_SAI\_Receive\_IT() &{SAI\_ReadData()} &  \\
          &       & HAL\_SAI\_Receive() &{SAI\_ReadMultiChannelBlocking()} &  \\
          &       &       &{SAI\_TransferReceiveNonBlocking()} &  \\
\cmidrule{3-5}          &       & HAL\_SAI\_Transmit\_DMA() &{SAI\_TransferSendNonBlocking()} &{HAL\_BE\_IO\_write()} \\
          &       & HAL\_SAI\_Transmit\_IT() &{SAI\_WriteBlocking()} &  \\
          &       & HAL\_SAI\_Transmit() &{SAI\_WriteMultiChannelBlocking()} &  \\
          &       &       &{SAI\_WriteData()} &  \\
\midrule
    \textbf{Storage} &{MMC} & HAL\_MMC\_Init() &{MMC\_Init()} &{HAL\_BE\_Storage\_Init()} \\
          &       & HAL\_MMC\_InitCard() &{MMC\_CardInit()} &  \\
\cmidrule{3-5}          &       & HAL\_MMC\_ReadBlocks() &{MMC\_ReadBlocks()} &{HAL\_BE\_Storage\_read()} \\
          &       & HAL\_MMC\_ReadBlocks\_IT() &{MMC\_ReadBootData()} &  \\
          &       & HAL\_MMC\_ReadBlocks\_DMA() &       &  \\
\cmidrule{3-5}          &       & HAL\_MMC\_WriteBlocks() &{MMC\_WriteBlocks()} &{HAL\_BE\_Storage\_write()} \\
          &       & HAL\_MMC\_WriteBlocks\_IT() &{MMC\_EraseGroups()} &  \\
          &       & HAL\_MMC\_WriteBlocks\_DMA() &       &  \\
          &       & HAL\_MMC\_Erase() &       &  \\
\cmidrule{2-5}          &{NAND} & HAL\_NAND\_Init() &       &{HAL\_BE\_Storage\_Init()} \\
\cmidrule{3-5}          &       & HAL\_NAND\_Read\_Page\_8b() &       &{HAL\_BE\_Storage\_read()} \\
          &       & HAL\_NAND\_Read\_SpareArea\_8b() &       &  \\
          &       & HAL\_NAND\_Read\_Page\_16b() &       &  \\
          &       & HAL\_NAND\_Read\_SpareArea\_16b() &       &  \\
\cmidrule{3-5}          &       & HAL\_NAND\_Write\_Page\_8b() &       &{HAL\_BE\_Storage\_write()} \\
          &       & HAL\_NAND\_Write\_SpareArea\_8b() &       &  \\
          &       & HAL\_NAND\_Write\_Page\_16b() &       &  \\
          &       & HAL\_NAND\_Write\_SpareArea\_16b() &       &  \\
          &       & HAL\_NAND\_Erase\_Block() &       &  \\
\cmidrule{2-5}          &{NOR} & HAL\_NOR\_Init() &       &{HAL\_BE\_Storage\_Init()} \\
\cmidrule{3-5}          &       & HAL\_NOR\_Read() &       &{HAL\_BE\_Storage\_read()} \\
          &       & HAL\_NOR\_ReadBuffer() &       &  \\
\cmidrule{3-5}          &       & HAL\_NOR\_Program() &       &{HAL\_BE\_Storage\_write()} \\
          &       & HAL\_NOR\_ProgramBuffer() &       &  \\
          &       & HAL\_NOR\_Erase\_Block() &       &  \\
          &       & HAL\_NOR\_Erase\_Chip() &       &  \\
\cmidrule{2-5}          &{SD} & HAL\_SD\_Init() &{SD\_Init()} &{HAL\_BE\_Storage\_Init()} \\
          &       & HAL\_SD\_InitCard() &{SD\_CardInit()} &  \\
\cmidrule{3-5}          &       & HAL\_SD\_ReadBlocks() &{SD\_ReadBlocks()} &{HAL\_BE\_Storage\_read()} \\
          &       & HAL\_SD\_ReadBlocks\_IT() &       &  \\
          &       & HAL\_SD\_ReadBlocks\_DMA() &       &  \\
\cmidrule{3-5}          &       & HAL\_SD\_WriteBlocks() &{SD\_WriteBlocks()} &{HAL\_BE\_Storage\_write()} \\
          &       & HAL\_SD\_WriteBlocks\_IT() &       &  \\
          &       & HAL\_SD\_WriteBlocks\_DMA() &       &  \\
          &       & HAL\_SD\_Erase() &       &  \\
\midrule
\textbf{Computing Accelerator} &{CRC} & HAL\_CRC\_Init() &{CRC\_Init()} &{HAL\_BE\_CRC\_config()} \\
          &       &       &{CRC\_WriteData()} &  \\
\cmidrule{3-5}          &       & HAL\_CRC\_Accumulate() &{CRC\_Get16bitResult()} &{HAL\_BE\_CRC\_cal\_result():} \\
          &       & HAL\_CRC\_Calculate() &{CRC\_Get32bitResult()} &  \\
\cmidrule{2-5}          &{CRYP} & HAL\_CRYP\_SetConfig() &{MMCAU\_AES\_SetKey} &{HAL\_BE\_CRYP\_config()} \\
\cmidrule{3-5}          &       & HAL\_CRYP\_Encrypt() &{mmcau\_AesCrypt()} &{HAL\_BE\_CRYP\_Enc()} \\
          &       & HAL\_CRYP\_Encrypt\_IT() &{mmcau\_DesCrypt()} &  \\
          &       & HAL\_CRYP\_Encrypt\_DMA() &       &  \\
\cmidrule{3-5}          &       & HAL\_CRYP\_Decrypt() &{mmcau\_AesCrypt()} &{HAL\_BE\_CRYP\_Dec()} \\
          &       & HAL\_CRYP\_Decrypt\_IT() &{mmcau\_DesCrypt()} &  \\
          &       & HAL\_CRYP\_Decrypt\_DMA() &       &  \\
\cmidrule{2-5}          &{HASH} & HAL\_HASH\_SHA1\_Accmlt()  &{MMCAU\_SHA1\_InitializeOutput()} &{HAL\_BE\_HASH\_sha1\_config()} \\
          &       & HAL\_HASH\_SHA1\_Accmlt\_IT() &       &  \\
          &       & HAL\_HASH\_SHA1\_Start\_DMA() &       &  \\
\cmidrule{3-5}          &       & HAL\_HASH\_SHA1\_Start() &{MMCAU\_SHA1\_HashN()} &{HAL\_BE\_HASH\_sha1\_get\_result()} \\
          &       & HAL\_HASH\_SHA1\_Accmlt\_End() &{MMCAU\_SHA1\_Update()} &  \\
          &       & HAL\_HASH\_SHA1\_Start\_IT() &       &  \\
          &       & HAL\_HASH\_SHA1\_Accmlt\_End\_IT() &       &  \\
          &       & HAL\_HASH\_SHA1\_Finish() &       &  \\
\cmidrule{3-5}          &       & HAL\_HASH\_MD5\_Accmlt() &{MMCAU\_MD5\_InitializeOutput()} &{HAL\_BE\_HASH\_md5\_config()} \\
          &       & HAL\_HASH\_MD5\_Accmlt\_IT() &       &  \\
          &       & HAL\_HASH\_MD5\_Start\_DMA() &       &  \\
\cmidrule{3-5}          &       & HAL\_HASH\_MD5\_Start() &{MMCAU\_MD5\_HashN()} &{HAL\_BE\_HASH\_md5\_get\_result()} \\
          &       & HAL\_HASH\_MD5\_Accmlt\_End() &{MMCAU\_MD5\_Update()} &  \\
          &       & HAL\_HASH\_MD5\_Start\_IT() &       &  \\
          &       & HAL\_HASH\_MD5\_Accmlt\_End\_IT() &       &  \\
          &       & HAL\_HASH\_MD5\_Finish() &       &  \\
\cmidrule{3-5}          &       &       &{MMCAU\_SHA256\_InitializeOutput()} &{HAL\_BE\_HASH\_sha256\_config()} \\
\cmidrule{3-5}          &       &       &{MMCAU\_SHA256\_HashN()} &{HAL\_BE\_HASH\_sha256\_get\_result()} \\
          &       &       &{MMCAU\_SHA256\_Update()} &  \\
\cmidrule{2-5}          &{PKA} & HAL\_PKA\_ModExp() &       &{HAL\_BE\_PKA\_mod\_config()} \\
          &       & HAL\_PKA\_ModExp\_IT() &       &  \\
          &       & HAL\_PKA\_ModExpFastMode() &       &  \\
          &       & HAL\_PKA\_ModExpFastMode\_IT() &       &  \\
\cmidrule{3-5}          &       & HAL\_PKA\_ModExp\_GetResult() &       &{HAL\_BE\_PKA\_mod\_get\_result()} \\
\cmidrule{3-5}          &       & HAL\_PKA\_RSACRTExp() &       &{HAL\_BE\_PKA\_crt\_config()} \\
          &       & HAL\_PKA\_RSACRTExp\_IT() &       &  \\
\cmidrule{3-5}          &       & HAL\_PKA\_RSACRTExp\_GetResult() &       &{HAL\_BE\_PKA\_crt\_get\_result()} \\
\cmidrule{3-5}          &       & HAL\_PKA\_PointCheck() &       &{HAL\_BE\_PKA\_ecc\_check\_config()} \\
          &       & HAL\_PKA\_PointCheck\_IT() &       &  \\
\cmidrule{3-5}          &       & HAL\_PKA\_PointCheck\_IsOnCurve() &       &{HAL\_BE\_PKA\_ecc\_check\_get\_result()} \\
\cmidrule{3-5}          &       & HAL\_PKA\_ECDSASign() &       &{HAL\_BE\_PKA\_ecdsa\_sign\_config()} \\
          &       & HAL\_PKA\_ECDSASign\_IT() &       &  \\
\cmidrule{3-5}          &       & HAL\_PKA\_ECDSASign\_GetResult() &       &{HAL\_BE\_PKA\_ecdsa\_sign\_get\_result()} \\
\cmidrule{3-5}          &       & HAL\_PKA\_ECDSAVerif() &       &{HAL\_BE\_PKA\_ecdsa\_verify\_config()} \\
          &       & HAL\_PKA\_ECDSAVerif\_IT() &       &  \\
\cmidrule{3-5}          &       & HAL\_PKA\_ECDSAVerif\_IsValidSignature() &       &{HAL\_BE\_PKA\_ecdsa\_verify\_config()} \\
\cmidrule{3-5}          &       & HAL\_PKA\_ECCMul() &       &{HAL\_BE\_PKA\_ecc\_sm\_config()} \\
          &       & HAL\_PKA\_ECCMul\_IT() &       &  \\
          &       & HAL\_PKA\_ECCMulFastMode() &       &  \\
          &       & HAL\_PKA\_ECCMulFastMode\_IT() &       &  \\
\cmidrule{3-5}          &       & HAL\_PKA\_ECCMul\_GetResult() &       &{HAL\_BE\_PKA\_ecc\_sm\_get\_result()} \\
\cmidrule{2-5}          &{RNG} & HAL\_RNG\_GenerateRandomNumber() &{RNGA\_GetRandomData()} &{HAL\_BE\_RNG\_get()} \\
          &       & HAL\_RNG\_GenerateRandomNumber\_IT() &       &  \\
          &       & HAL\_RNG\_ReadLastRandomNumber() &       &  \\
\label{peripheral_drivers}
\end{longtable}
} 
\clearpage
\twocolumn